\documentclass{aastex6}
\usepackage{color}
\definecolor{Cafe}{rgb}{0.48, 0.25, 0.0}
\definecolor{v}{rgb}{0.55, 0.0, 0.55} 

\usepackage{soul}

\begin{document}


\title{Strong lensing modeling in galaxy clusters as a promising method to test
cosmography I. Parametric dark energy models}


\author{Juan Maga\~na\altaffilmark{1}}
\affil{Instituto de F\'{\i}sica y Astronom\'ia, Facultad
de Ciencias\\ 
Universidad de Valpara\'iso, Avda. Gran Breta\~na 1111,\\
Valpara\'iso, Chile. \\
}

\author{Ana Acebr\'on\altaffilmark{2}}
\affil{Aix Marseille Univ, CNRS, CNES, LAM, Marseille, France}
\affil{Physics Department, Ben-Gurion University of the Negev, P.O. Box 653, Be'er-Sheva 8410501, Israel}

\author{Ver\'onica Motta\altaffilmark{3}}
\affil{Instituto de F\'{\i}sica y Astronom\'ia, Facultad
de Ciencias\\ 
Universidad de Valpara\'iso, Avda. Gran Breta\~na 1111,\\
Valpara\'iso, Chile. \\
}

\author{Tom\'as Verdugo\altaffilmark{4}}
\affil{Instituto de Astronom\'ia, Universidad Nacional Aut\'onoma de M\'exico, Apartado postal 106, C.P. 22800,  Ensenada, B.C., M\'exico}

\author{Eric Jullo\altaffilmark{5}}
\affil{Aix Marseille Univ, CNRS, CNES, LAM, Marseille, France}

\and
\author{Marceau Limousin\altaffilmark{6}}
\affil{Aix Marseille Univ, CNRS, CNES, LAM, Marseille, France}

\altaffiltext{1}{juan.magana@uv.cl}

\begin{abstract}

In this paper we probe five cosmological models for which the dark energy equation
of state parameter, $w(z)$, is parameterized as a function of redshift using
strong lensing data in the galaxy cluster Abell 1689. We constrain the 
parameters of the $w(z)$ functions by reconstructing the lens model under each one
of these cosmologies with strong lensing measurements from two galaxy clusters: \textit{Abell 1689} and a mock cluster, \textit{Ares}, from the Hubble Frontier Fields Comparison Challenge, to validate our methodology. To quantify how the cosmological constraints are biased due to systematic effects in the strong lensing modeling, we carry out three runs considering the following uncertainties for the multiple images positions: $0.25\arcsec$, $0.5\arcsec$, and $1.0\arcsec$. With \textit{Ares}, we find that larger errors decrease the systematic bias on the estimated cosmological parameters. With real data, our strong-lensing constraints on $w(z)$ are consistent those derived from other cosmological probes.
We confirm that strong lensing cosmography with galaxy clusters is a promising method to constrain $w(z)$ parameterizations. 
A better understanding of galaxy clusters and their environment is however needed to improve the SL modeling and hence to estimate stringent cosmological parameters in alternatives cosmologies.

\end{abstract}

\keywords{dark energy, strong lensing, galaxy clusters}

\section{Introduction} \label{sec:intro}
Current cosmological observations provide strong evidence that the expansion of the Universe is accelerating \citep{Riess:1998,Perlmutter:1999,Planck:2015XIV}. 
The source of this cosmic acceleration is a big puzzle in modern cosmology and two hypothesis have been proposed to explain it: either to postulate the existence 
of a dark energy component or to modify the gravity laws \citep{Joyce:2016}.
Among the first kind of models, the cosmological constant, which is commonly associated to the quantum vacuum energy, 
has been established as the preferred candidate to the nature of dark energy by several cosmological measurements \citep[e.g.][]{Planck:2015XIII}. By definition, the equation of state (EoS, hereafter) parameter of the cosmological constant is $w=-1$. Nonetheless, when a general constant equation of state is considered, the data constrain $w=-1.019^{+0.075}_{-0.080}$ \citep[][see also \citealp{Neveu:2017}]{Planck:2015XIV}, which is consistent with the cosmological constant.
In spite of this consistency, the theoretical expected value of the vacuum energy differs in many orders of magnitude from the observed one. In addition, the coincidence problem, i.e. the similitude seen at the current time between the dark matter energy density and that of DE, remains unsolved \citep{Zeldovich,Weinberg}.
Several dark energy (DE, hereafter) models, as for instance, dynamical dark energy or interacting dark energy  \citep{Copeland:2006,Li:2011},
are also in agreement with the data and they can satisfactorily describe the late-time acceleration of the Universe in a similar way as the cosmological constant does \citep{Ferreira:2017, Salvatelli:2014, Zhao:2017Nat}. 
Therefore, to distinguish which cosmological model is the more suitable to the nature of dark energy, we need to put tight constraints on their parameters. 
A standard way to estimate these parameters is to perform a Bayesian analysis using classic cosmological probes, i.e. to fit the distance modulus
of type Ia distant supernovae (SNIa), Hubble parameter measurements, baryon acoustic oscillation (BAO) signal, 
and the acoustic peaks of the cosmic microwave background (CMB) radiation \citep{Davis:2014,Mortonson:2014}. 
Although these tests are widely used to constrain cosmological models, they could yield to biased estimations because either the data or the test fitting formulas are derived assuming an
underlying standard cosmology $\Lambda\mathrm{CDM}$ (i.e. the cosmological constant as dark energy plus cold dark matter).
Thus, it is essential to construct methods to estimate the parameters of alternative cosmologies without assuming any fiducial cosmology. One novel technique is to use strong lensing measurements in galaxy clusters.\\

Strong gravitational lensing (SL, hereafter) offers a unique and independent opportunity to constrain dark energy features without prior assumptions on the fiducial cosmology. \citet{Link:1998ApJ} introduced a new approach by leveraging the cosmological sensitivity of the angular size-redshift relation when multiples imaged systems (over a broad range of redshift) are produced by strong lensing clusters.
This technique was later on extended to more complex simulated clusters by \citet{Golse:2002} and to real clusters such as Abell 2218 \citep[see][]{Soucail:2004}, showing that SL cosmography is a promising geometrical cosmological test.
\citet{Jullo:2010} used an improved technique which simultaneously reconstructed the mass distribution of Abell 1689 (A1689, hereafter), adopting a parametric lens modeling, and constrained the parameters of a $w$CDM cosmology.
For the first time, the authors obtained competitive constraints on the equation of state parameter and found that, by combining their results with other probes, they improved the DE EoS estimation by $\sim30\%$. Following the same method,
\citet{Caminha:2016} recently used the SL measurements in Abell S1063 with the pre-\textit{Frontier Fields} data to constrain cosmological parameters for three different $\Lambda\mathrm{CDM}$ models. They pointed out the importance of estimating the parameters using multiply lensed sources with a wide range of redshifts. The authors also showed that the lack of spectroscopic measurements or the use of inaccurate photometric redshifts leads to a biased estimation of the cosmological parameters. 
\citet{Magana:2015} exploited this technique too, but using alternative cosmologies. They used A1689 strong lensing measurements to constrain four dark energy models: Chevallier-Polarski-Linder (CPL), Interacting Dark Energy (IDE), Ricci Holographic Dark Energy (RHDE), and Modified Polytropic Cardassian (MPC) . 
They found that the SL method provides CPL constraints in good agreement with those obtained with the SNe Ia, BAO and CMB data. In addition, the IDE and RHDE constraints derived from SL are similar to those estimated with other tests. 
Nevertheless, the IDE constraints are consistent with the complementary bounds only if an increase in the image-position error \citep[five times the one previously used by][]{Jullo:2010}
is considered in the lens modeling.
They confirmed that, to avoid misleading DE bounds, it is important to consider larger positional uncertainties for the multiple images; which could be associated with systematic errors.\\

Indeed, SL has various known sources of systematic errors. \citet{D'Aloisio:2011}, using simulations of cluster lenses, showed that the observational errors (for space-based images) are an order of magnitude smaller that the modeling errors. Furthermore, line-of-sight (LOS) structures can introduce a systematic error in the strong lensing modeling \citep[e.g.][]{Bayliss:2014, Giocoli2016,Host:2012, Jaroszynski:2014, McCully:2014} of up to $\sim 1.4\arcsec$ on the position of multiple images \citep{Zitrin2015}. Even distant massive structures in the lens plane have a significant impact on the position of multiple images \citep{Tu2008, Limousin2010}. 
\citet{Harvey2016}, by analyzing the \textit{Frontier Field} cluster MACSJ0416 ($z = 0.397$), estimated an error of $\sim 0.5\arcsec$ on the position of the multiple images when assuming that light traces mass in the SL modeling.
However, few studies have investigated their impact on the retrieval of cosmological parameters \citep{McCully2017, Acebron2017}. \\

In this paper, we are interested in quantifying the uncertainties in the estimation of cosmological parameters induced by different positional errors of the multiple images. To this end, we analyze the strong lensing effect in the galaxy cluster A1689, as well as in a mock galaxy cluster at $z=0.5$ generated in a flat ${\Lambda\mathrm{CDM}}$ cosmology. Because in the CLP case it is possible to obtain tight constraints on its parameters \citep[see][]{Magana:2015} using the SL methodology proposed by \citet{Jullo:2010}, in this work we consider popular CPL-like models in which the EoS of dark energy is parametrized as function of redshift.\\ 

The paper is organized as follows: in the next section, \S \ref{sec:CosmoFrame}, we introduce the cosmological framework and the parametric dark energy models. 
In section \S \ref{sec:SLmethod} we describe the SL data and methodology used to constrain the cosmological parameters of the DE models. 
In section \S \ref{sec:results} we present and discuss the results. Finally, we provide our conclusions in section \S \ref{sec:conclusions}.

\section{Cosmological framework and parametric dark energy models}\label{sec:CosmoFrame}
For a homogeneous, isotropic, and flat Friedmann-Lema\^itre-Robertson-Walker (FLRW) cosmology, the expansion rate of the Universe is governed by the Friedmann equation:
\begin{equation}
H^{2}(z)\equiv \frac{8\pi G}{3}\sum_{i} \rho_{i}(z),
\label{eq:gFriedmann}
\end{equation}
where $H\equiv \dot{a}/a$ is the Hubble parameter, $a$ is the scale factor of the Universe, and
$\rho_{i}$ denotes the energy density for each component in the Universe\footnote{dot stands for the derivative with respect to the cosmic time}.
We consider cold dark matter ($m$) and radiation ($r$) components
whose dynamics are described by a perfect fluid with EoS $w_{m}=0$ and $w_{r}=1/3$, respectively.
In addition, we also consider a dynamical dark energy ($de$) whose EoS is parameterized by a 
$w(z)$ function. In terms of the present values\footnote{Quantities evaluated at $z=0$} of the density parameters, 
$\Omega_{i} \equiv 8 \pi G \rho_{i}/3 H(z)^{2}$, for each component, the Eq. \ref{eq:gFriedmann} reads as:
\begin{equation}
 E^{2}(z)=\Omega_{m}(1+z)^{3} + \Omega_{r}(1+z)^{4} +\Omega_{de}f_{de}(z),\quad
\label{eq:Ez}
\end{equation}
\noindent where $E(z)=H(z)/H_{0}$ is the dimensionless Hubble parameter, $\Omega_{r}=2.469\times10^{-5}h^{-2}(1+0.2271 N_{eff})$, 
with $h=H_{0}/100\mathrm{kms}^{-1}\mathrm{Mpc}^{-1}$,  $N_{eff}=3.04$ is the standard number of relativistic species \citep{Komatsu:2011}, 
and $\Omega_{de}$ can be expressed as $\Omega_{de}=1-\Omega_{m}-\Omega_{r}$. The function $f_{de}(z)$ is defined as:
\begin{equation}
f_{de}(z)\equiv \frac{\rho_{de}(z)}{\rho_{de}(0)}=
\mathrm{exp}\left(3\int^{z}_{0}\frac{1+w(z)}{1+z}\mathrm{dz}\right).
\label{eq:xz}
\end{equation}
\noindent
Notice that, by introducing a $w(z)$ functional form in the integral of the Eq. (\ref{eq:xz}),
we can obtain an analytical expression for $f_{de}(z)$, and hence for $E(z)$.

Besides, to test whether the constraints for each parametric DE model result in a late cosmic acceleration, we examine the deceleration parameter \textit{q(z)} defined as:
\begin{equation}
q(z) = - \frac{\ddot{a}(z)a(z)}{\dot{a}^{2}(z)}.
\label{eq:qa}
\end{equation}
\noindent
Using Eq.\,(\ref{eq:Ez}), we obtain:

\begin{equation}
q(z) = \frac{(1+z)}{E(z)} \frac{dE(z)}{dz}-1,
\label{eq:qz}
\end{equation}
\noindent
which expresses the deceleration parameter in terms of the dimensionless Hubble parameter.

\subsection{Parametric dark energy models}\label{sec:DEmod}
One alternative to the cosmological constant is to consider a
dark energy component which admits a time-dependent EoS. 
 An effective and simple way to study dynamical dark energy models 
is to assume a phenomenological parameterization of the EoS \citep{Lazkoz:2005,Pantazis:2016}. Commonly, 
this EoS is biparametric and it depends on the scale factor of redshift.
The most popular ansatz, denoted Chevallier-Polarski-Linder parameterization, \citep[introduced and revisited by][respectively]{CP:2001, Linder:2003} 
is $w(z)=w_{0}+w_{1} z/(1+z)$, where $w_{0}$ is the present value of the equation of state
and $w_{1}=d w(z)/ dz|_{z=0}$. In this paper we study five CPL-like EoS parameterizations \citep[see][for details]{Magana:2014,Wang:2016}, in the following, we briefly introduce the functional form of these parameterizations.

\begin{itemize}
 \item Jassal-Bagla-Padmanabhan (JBP).- \citet{Jassal:2005a, Jassal:2005b} proposed that the dark energy EoS is parameterized
 by the function
\begin{equation}
w(z)=w_{0} + w_{1}\frac{z}{\left(1+z\right)^{2}},
\label{eq:wJBP}
\end{equation}
which allows rapid variations at low $z$. The DE has the same EoS at the present epoch and at high
redshift, i.e., $w(\infty)=w_{0}$. By substituting the Eq. (\ref{eq:wJBP}) in Eq. (\ref{eq:xz}) we obtain:
\begin{equation}
f_{de}(z)=(1+z)^{3(1+w_{0})}\mathrm{exp}\left[\frac{3}{2}\frac{w_{1}z^{2}}{(1+z)^{2}}\right].
\label{eq:fzJBP}
\end{equation}

\item Barbosa-Alcaniz (BA).- \citet{Barboza:2008} considered a parametric EoS for the dark energy component
given by:
\begin{equation}
w(z)=w_0 + w_1 \frac{z(1+z)}{1+z^2}.
\label{eq:wBA}
\end{equation}
This ansatz behaves linearly at low redshifts as $w_{0}+w_{1}$, and $w\rightarrow w_{0}+w_{1}z$
when $z\rightarrow\infty$. In addition, $w(z)$ is well-behaved in 
all epochs of the Universe, for instance, the DE dynamics in the future, at $z=-1$, can be investigated
without dealing with a divergence. Solving the integral in Eq. (\ref{eq:xz}) and using Eq. (\ref{eq:wBA}) results in:

\begin{equation}
f_{de}(z)=(1+z)^{3(1+w_0)}(1+z^2)^{\frac{3}{2}w_{1}}. 
\label{eq:fzBA}
\end{equation}

\item Feng-Shen-Li-Li \citep[FSLL,][]{Feng:2012} suggested two dark energy EoS parameterizations given by:
\begin{equation}
w(z)=w_{0} + w_{1}\frac{z}{1+z^{2}}, \qquad \textrm{FSLLI}
\label{eq:wFSLLI}
\end{equation}

\begin{equation}
w(z)=w_{0} + w_{1}\frac{z^{2}}{1+z^{2}} \qquad \textrm{FSLLII}.
\label{eq:wFSLLII}
\end{equation}
\noindent
Both functions have the advantage of being divergence-free throughout the entire cosmic evolution, even at $z=-1$. At low redshifts, 
$w(z)$ behaves as $w_{0}+w_{1}z$ and $w_{0}+w_{1}z^{2}$ for FSLLI and FSLLII respectively. In addition,
when $z\rightarrow \infty$, the EoS has the same value, $w_{0}$, as the present epoch for FSLLI and $w_{0}+w_{1}$ for FSLLII.
Using Eqs. (\ref{eq:wFSLLI})-(\ref{eq:wFSLLII}) to solve Eq. (\ref{eq:xz}) leads to:

\begin{equation}
f_{de\pm}(z)=(1+z)^{3(1+w_{0})}\mathrm{exp}\left[\pm\frac{3w_{1}}{2}\mathrm{arctan(z)}\right]
\left(1+z^{2}\right)^{\frac{3}{4}w_{1}}\left(1+z\right)^{\mp \frac{3}{2}w_{1}},
\label{eq:fzFSLL}
\end{equation}
\noindent
where $f_{+}$ and $f_{-}$ correspond to FSLLI and FSLLII respectively.

\item Sendra-Lazkoz \citep[SeLa,][]{Sendra:2012}  improved the CPL parameterization, whose
$w_{0}-w_{1}$ parameters are highly correlated and $w_{1}$ is poorly constrained by the observational data,
introducing new polynomial parameterizations. They are constructed to reduce the parameter correlation,
so they can be better constrained by the observations at low redshifts. One of these parameterizations is given by:

\begin{equation}
w(z)=-1 + c_{1}\left(\frac{1+2z}{1+z}\right) + c_{2}\left(\frac{1+2z}{1+z}\right)^{2},\\ 
\label{eq:wSeLa}
\end{equation}
\noindent
where the constants are defined as $c_{1}=(16w_{0}-9w_{0.5}+7)/4$, and $c_{2}=-3w_{0}+ (9w_{0.5}-3)/4$,
and $w_{0.5}$ is the value of the EoS at $z=0.5$.
This $w(z)$ function is well-behaved at higher redshifts as $(-1-8w_{0}+9w_{0.5})/2$. 
By the substitution of Eq. (\ref{eq:wSeLa}) into Eq. (\ref{eq:xz}), we obtain:

\begin{equation}
f_{de}(z)=(1+z)^{\frac{3}{2}(1-8w_{0}+9w_{0.5})}
\mathrm{exp}\left[\frac{3z\left\{w_{0}(52z+40)-9w_{0.5}(5z+4)+7z+4\right\}}{8(1+z)^2}\right]. 
\label{eq:fzSeLa}
\end{equation}
\end{itemize}

By replacing the $f_{de}(z)$ functions in Eq. (\ref{eq:gFriedmann}), we obtain an analytical $E(z)$ function for each parametric $w(z)$, which will be used in the following sections to estimate the EoS parameters. Our main
purpose is to examine the quality of the $w(z)$ constraints extracted from the SL modeling when different image-position errors are considered.

\section{Methodology} \label{sec:SLmethod}

\subsection{Strong lensing as a cosmological probe}

The gravitational lensing effect is produced when the light-beam of a background source is deflected by a gravitational lens, i.e. a mass distribution between the source and the observer. We refer to the strong lensing regime when several rings, arcs or multiples images are observed as a result of the distortion and deflection of the light from a source by a lens. These strong lensing observables offer a powerful and useful tool to not only infer the total matter distribution in astrophysical systems \citep{Jauzac2014,Monna2017}, but also to provide insights on the total content of the Universe, dark matter and dark energy properties \citep{Golse:2002,Soucail:2004, Jullo:2010,Caminha:2016, Magana:2015}. Here, we use strong lensing measurements in galaxy clusters to constrain the equation of state of parametric dark energy models.\\

Since the strong lensing features depend on the dynamics of the Universe via the angular diameter distance between the source, lens and observer, it can be used as a geometric cosmological probe.
For any underlying cosmology, the angular diameter distance ratios for two images from different sources defines the 'family ratio' \citep[see][for a detailed discussion]{Jullo:2010}:

\begin{equation}
\Xi(z_1,z_{s1},z_{s2},\mathbf{\Theta}) = \frac{D(z_1,z_{s1})}{D(0,z_{s1})}\frac{D(0,z_{s2})}{D(z_1,z_{s2})},
\label{eq:Xi}
\end{equation}

\noindent where $\mathbf{\Theta}$ is the vector of cosmological parameters to be fitted,
$z_1$ is the lens redshift, $z_{s1}$ and $z_{s2}$
are the two source redshifts, and $D(z_i,z_f)$ is the angular
diameter distance calculated as:

\begin{equation}
D(z_{i},z_{f}) =  \frac{r(z_i,z_f)}{(1+z_f)},
\label{eq:dA}
\end{equation}
\noindent
where $r(z_i,z_f)$, the comoving distance of a source at redshift $z_f$ measured by an observer at redshift $z_i$, is given by

\begin{equation}
r(z_i,z_f)=\frac{c}{H_0}\int_{z_i}^{z_f} \frac{dz'}{E(z')}.
\label{eq:rz}
\end{equation}

Notice that 
the underlying cosmology in the lens modeling is selected by introducing the $E(z)$ function
in the Eq. (\ref{eq:rz}). For the parametric DE models, these functions are analytical and 
$\mathbf{\Theta}=\{\Omega_{m}, w_{0}, w_{1}\}$ ($w_{0.5}$ for the SeLa parameterization) is the free parameter vector. 

\subsection{Lensing modeling}

To constrain the parameters of the DE models presented in \ref{sec:DEmod}, we use the SL measurements in two galaxy clusters: a real one, \textit{Abell 1689}, and a simulated one, \textit{Ares} from the Frontier Fields Comparison Challenge \citep{meneghetti2016}. \\
We performed the SL modeling using the public software LENSTOOL\footnote{\url{https://projets.lam.fr/projects/lenstool}} \citep{kneib1996, Jullo:2007} 
in which the DE cosmological models described in \ref{sec:DEmod} were implemented. LENSTOOL is a ray-tracing code with
a Bayesian Markov Chain Monte-Carlo sampler which optimizes the model parameters using the positions of the multiply imaged systems. 
The matter distribution in clusters is modeled in a parametric way and the optimization is performed in the image plane for \textit{Abell 1689} as it is more precise
\citep[this is different from the analysis by][where the optimization was performed in the source plane]{Jullo:2010,Magana:2015}. 
For \textit{Ares}, the optimization was realized in the source plane as it is a more complex cluster (more images and cluster members) and this procedure is more computationally efficient. We checked that results in the image plane were similar for a subset of calculations.\\

For both \textit{Abell 1689} and \textit{Ares}, each potential (either large or galaxy-scale) is parametrized with the Pseudo Isothermal Elliptical Mass Distribution profile \citep[hereafter PIEMD,][]{kassiola1993,Eliasdottir:2007}. The density distribution of this profile is given by: 

\begin{equation}
\centering
\rho(r) = \frac{\rho_{0}}{(1 + \frac{r^{2}}{{r_{core}}^{2}})(1 + \frac{r^{2}}{{r_{cut}}^{2}})}\mathrm{,}
\label{eq:piemd}
\end{equation}

\noindent with a central density $\rho_{0}$, a core radius $r_{core}$ and a truncation radius $r_{cut}$. This profile is characterized by two changes in the density slope: it behaves as an isothermal profile within the transition region but the density falls as $\rho \propto r^{-4}$ at large radii. In LENSTOOL, it has the following free parameters: the coordinates x, y; the ellipticity, e; angle position, $\theta$; core and cut radii, $r_{core}$ and $r_{cut}$ and a velocity dispersion, $\sigma$.
Both clusters were modeled in the same way regardless of the considered DE cosmological model.\\

\textit{Abell 1689.-} a massive cluster at redshift $z=0.18$, is one of the most studied strong lenses \citep[see e.g.,][and references therein]{Limousin:2007ApJ,Limousin2013,Umetsu2015,Diego2015}. The first SL modeling was performed by \citet{Miralda1995} which already required a bi-modal mass distribution for the cluster. It is one of the most X-ray luminous clusters and has a large Einstein radii, $\sim 45\arcsec$.
A1689 is still the target of recent observations, using MUSE data with which \citet{Bina2016} confirmed or spectroscopically identified new multiples images as well as cluster members.\\

We refer the reader to \citet{Jullo:2010} for a detailed discussion of the modeling of A1689, where a SL parametric model was used to constrain the DE EoS. As we follow-up their approach, we give here a quick overview. 
\textit{Abell 1689} was modeled using the SL features in the deep HST observations and extensive ground-based spectroscopic follow-up. The mass distribution was represented as bi-modal, with one central and dominant large-scale potential harboring a brightest central cluster galaxy (BCG) in its centre. The second large-scale potential was situated in the north-east. \citet{Jullo:2010} used 58 cluster galaxies (with $m_K<18.11$) in the modeling and followed the standard scaling relations. In this work we consider the same catalog as in \citet{Jullo:2010}, including 28 images from 12 families\footnote{a family is the group of images associated to one lensed source}, all with measured spectroscopic redshifts, spanning a range of 1.15 $< z_{S} <$ 4.86.\\


\textit{Ares.-} a mock galaxy cluster at $z=0.5$, generated in a flat ${\Lambda\mathrm{CDM}}$ cosmological model with a matter density parameter $\mathrm{\Omega_{m}}=0.272$. We model \textit{Ares} considering all multiple images (242 from 85 sources), all with assumed known spectroscopic redshifts spanning a wide range ($0.91 < z_{S} < 6.0$). Cluster members are taken from the given simulated catalogue up to a magnitude of $m_{F160W} < 22.0$~mag (representing $> 90\%$ of the total cluster luminosity). \textit{Ares} is part of an archive of mock clusters which reproduce the characteristics of the \textit{Frontier Fields} observations \citep[the FF-SIMS Challenge,][]{meneghetti2016}. It was part of a challenge among the strong lensing community to perform, first a blind reconstruction of the mass distribution of the cluster, and then to improve the models after the unblinding of the true mass distribution. The conclusions of this challenge were primarily used to calibrate different modeling techniques. 
\textit{Ares} is a semi-analytical cluster created with MOKA\footnote{\url{https://cgiocoli.wordpress.com/research-interests/moka/}} by \citet{giocoli2012}. This simulated cluster, a bimodal and realistic cluster, is built with three components: two smooth dark matter triaxial haloes, two bright central cluster galaxies (BCGs) and a large number of sub-haloes. Dark matter sub-haloes are populated using a Halo Occupation Distribution technique (HOD) and stellar and B-band luminosities are given for all galaxies according to the mass of their sub-halo as in \citet{wang2006}.\\
We have modeled \textit{Ares} as two large-scale potentials and two potentials for the BCGs, whose coordinates are fixed as well as the ones for the large-scale potentials \citep[see Figure 1 in][]{Acebron2017}. Both components have been parametrized with the PIEMD density profile and corresponds to the model PIEMD - PIEMD in \citet{Acebron2017}.
The modeling also includes cluster galaxies with $m_{F160W} < 22$~mag (being a more complex cluster, computing time is reduced by introducing  a magnitude cut representing $>90\%$ of the total cluster luminosity) with masses scaling with luminosity \citep[see][for further details]{limousin2005}. Three massive cluster galaxies close to multiple images \citep[see Figure 1 in][]{Acebron2017} were more carefully modeled (i.e their parameters deviate from the scaling relations). All multiple images provided were taken into account in the modeling, resulting in an average positional accuracy of $0.66\arcsec$ and giving tight constraints on the $\Omega_{M} - w$ space parameter considering a flat ${\Lambda\mathrm{CDM}}$ cosmology .\\

To quantify how the cosmological constraints are biased due to systematic effects in the SL modeling, we use different image-positional errors, $\delta_{pos}$, and compare the resulting DE parameter estimations.  
For each parametric DE model and for both \textit{Abell 1689} and \textit{Ares}, 
we carry out three runs considering the following errors for the multiple images positions: $0.25\arcsec$, $0.5\arcsec$, and $1.0\arcsec$. These values are chosen arbitrarily but they intend to cover the range of values of systematic uncertainties reported by different authors \citep{Zitrin2015, Harvey2016}. For instance, the observations indicate uncertainties on the image positions $\sim0.06\arcsec$ \citep[][see also \citealt{Chirivi:2018}]{Grillo:2015ApJ} which is almost one order of magnitude less that our smaller error. Nevertheless, the same authors increase this error in the modeling up to six times, i.e $0.4\arcsec$, to take into account systematics due to the LOS structures and small dark-matter clumps. Although an error of $0.5\arcsec$ is in agreement with predictions of the effects of matter density fluctuations
along the LOS \citep{Host:2012,Caminha:2016}, other authors claim that $1.4\arcsec$  is the reasonable error to account for these systematics in lens modeling \citep[e.g.][]{Zitrin:2012}. Finally, \citet{Chirivi:2018}  proposed to use different errors in the range $0.2\arcsec$-$0.4\arcsec$ for different images. In order to consider all these possible effects, we propose $0.25\arcsec$ as minimum error in the position of images and increase it two times in each run.

The best-fitting model parameters are found by minimizing the distance between the observed and model-predicted positions of the multiple images.
To assess the goodness of the lens model fit we examine the reduced chi-square $\chi^{2}_{red}$ \citep[see][for details how it is calculated in the source and image plane]{Jullo:2007}. We also use the root-mean-square between the observed and predicted positions of the multiple images from the modeling, computed as follows: 
\begin{equation}
RMS = \sqrt{\frac{1}{N} \sum_{i=1}^{n} |\theta_i^{obs} - \theta_i^{pred}|^2}\mathrm{,}
\end{equation}
where $\theta_i^{obs}$ and $\theta_i^{pred}$ are the observed and model-predicted positions of the multiples images and N being the total number of images. Although both estimators are widely used to compare the goodness-of-fit of the cluster parameters among different lens models \citep[e.g.][]{Jullo:2010,Limousin:2016,Caminha:2016}, they are less sensitive to the cosmological parameters. A reliable tool to measure the goodness of fit for the cosmological constraints is the Figure-of-Merit \citep[FOM,][]{Wang:2008} given by
\begin{equation}
\mbox{FoM}=\frac{1}{\sqrt{\det \mbox{Cov} (f1,f2,f3,...)}},
\label{eq:fom}
\end{equation}
\noindent
where $\mbox{Cov} (f1,f2,f3,...)$ is the covariance matrix of the
cosmological parameters {$f_{i}$}. This indicator (Eq. \ref{eq:fom}) is a generalization of those proposed by \citet{Albrecht:2006} and larger values imply stronger constraints on the cosmological parameters since it corresponds to a smaller error ellipse.

\section{Results}\label{sec:results}
In this section we present, for each cosmological model, the constraints from the strong lensing measurements of \textit{Ares} and \textit{Abell 1689}, as well as those from other complementary probes (H(z), SNe Ia, BAO, and CMB, see Appendix \ref{sec:Appendix}). The mock \textit{Ares} cluster has the advantage of being able to directly compare and validate the cosmological constraints from the SL technique with the fiducial cosmology i.e. the $\Lambda$CDM with $\Omega_{m}=0.272$.\\
For each cosmological model, the cluster model parameters and the cosmological parameters are simultaneously optimized with the \textsc{LENSTOOL} software with 80000 MCMC steps. For the complementary probes, we carry out the \textsc{EMCEE} python module \citep{Foreman-Mackey:2013} employing 500 walkers, 2500 burn-in phase steps, and 7500 MCMC steps to guaranty the convergence. In all our estimations, we have adopted a dimensionless Hubble constant $h=0.70$.\\
Tables \ref{tabJBP}-\ref{tab:SeLa} show the mean parameters obtained after optimization for both \textit{Abell 1689}, \textit{Ares} and each cosmological model (JBP, BA, FSLL I, FSLL II, and SeLa). For each positional uncertainty considered, we present the $\chi^2_{red}$, the RMS in the image plane as well as the mean values obtained for the cosmological parameters fitted, $\Omega_{m}$, $w_{0}$ and $w_{1}$ with the $68\%$ uncertainties. The same Tables also give the mean values for the DE parameters obtained from the complementary probes. The best fit and $2D$ confidence contours for the cosmological parameters were computed using the python module Getdist\footnote{it can be download in https://github.com/cmbant/getdist}.

\subsection{Effect of image-position error on the cosmological parameters}

The positional error for the multiple images plays a key role in both the lens modeling and the cosmological parameter estimation \citep{Limousin:2016,Magana:2015,Caminha:2016}. 

\begin{figure}
\figurenum{1}
\gridline{\fig{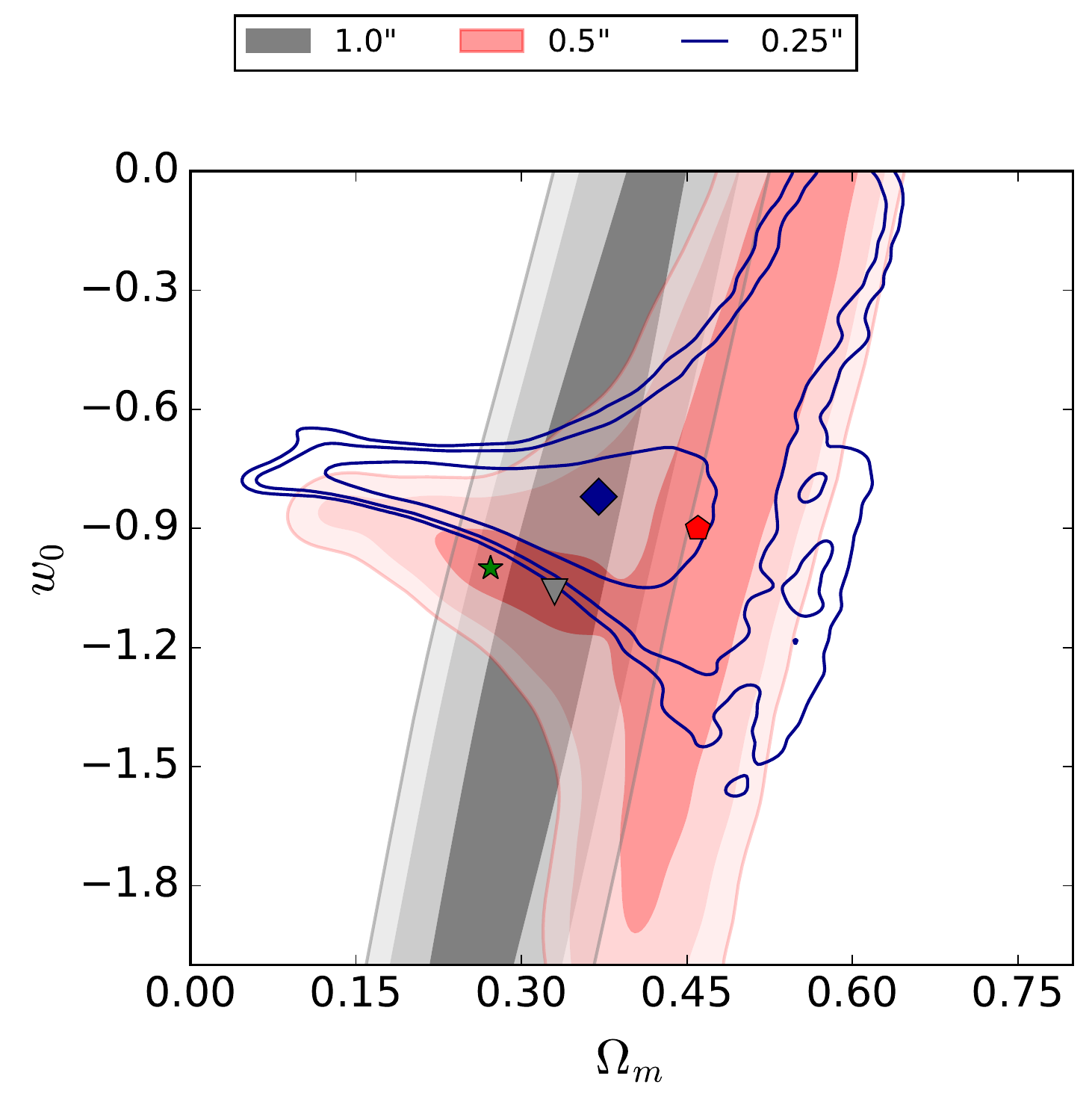}{0.45\textwidth}{}
          \fig{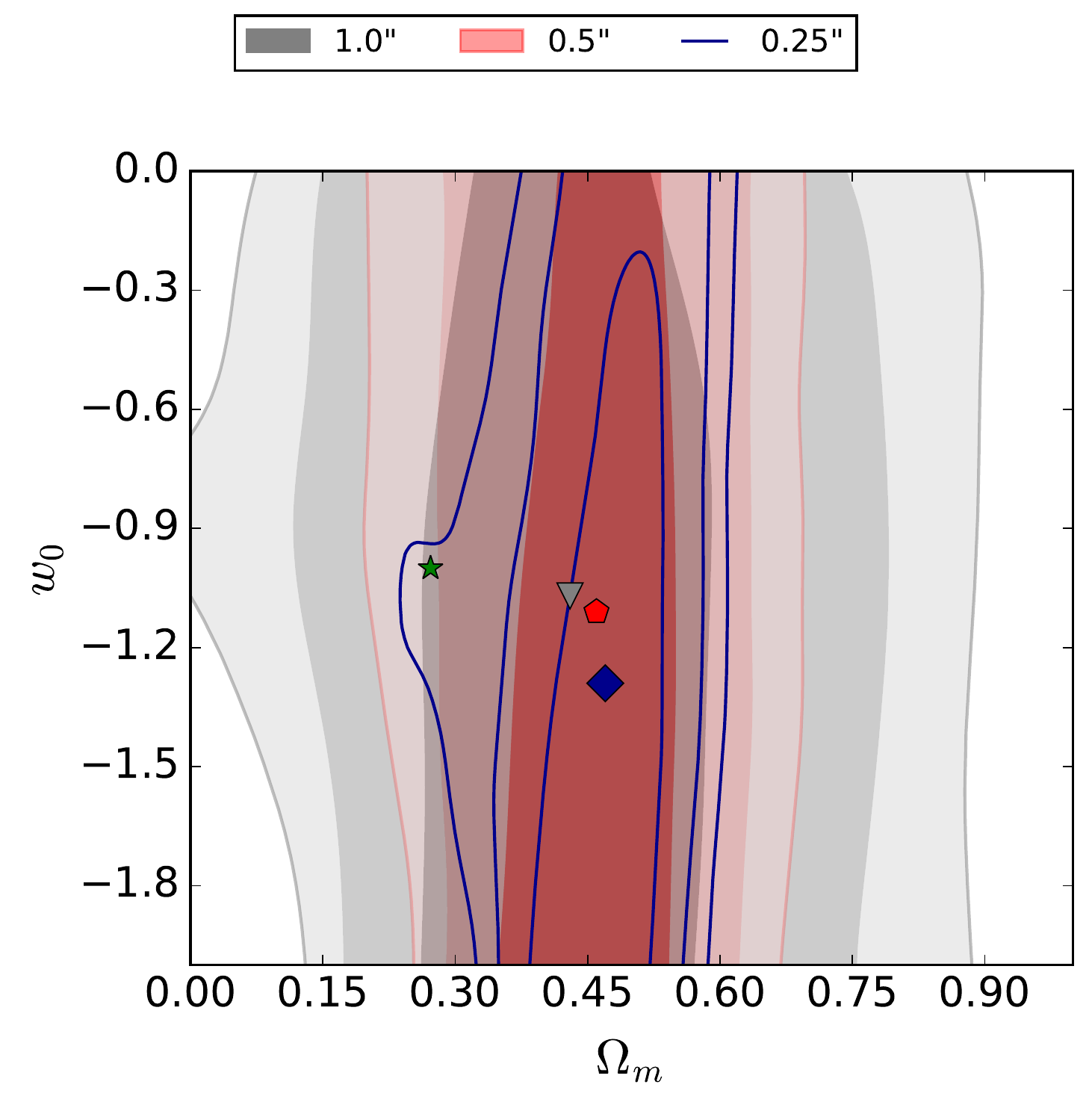}{0.45\textwidth}{}
          }
\caption{Comparison of the constraints on the $\mathrm{\Omega_m}$-$w_0$ parameters for \textit{Ares} (left panel) and \textit{Abell1689} (right panel) when considering several positional uncertainties ($\delta_{pos}$) in the SL modeling for the JBP parameters. The star indicates the reference fiducial values. The mean values when $\delta_{pos}$ is $0.25\arcsec$, $0.5\arcsec$, and $1.0\arcsec$ are represented by the diamond, pentagon and triangle respectively.}
\label{fig:JBPmodelcompare}
\end{figure}

As mentioned in \S\ref{sec:intro}, a large error could take into account other sources of uncertainties in the SL measurements, such as systematic errors due to foreground and background structures \citep{D'Aloisio:2011,Host:2012,Bayliss:2014,Zitrin2015} or the cluster's environment \citep{McCully2017, Acebron2017}. 
As a first test, we constrain the JBP cosmological parameters from \textit{Ares} SL data with an image positional error $\delta_{pos}=1\arcsec$. For this run (see Table \ref{tabJBP}), we obtain $\chi^{2}_{red}=0.77$ and $\mathrm{RMS}=0.94\arcsec$, both criteria indicating a good-fit for the cluster parameters.  When modeling \textit{Ares} using smaller positional uncertainties ($\delta_{pos}=0.25\arcsec, 0.5\arcsec$), the $\chi^{2}_{red}$ values point out a poor model fit.
The left panel of Figure \ref{fig:JBPmodelcompare} illustrates the comparison of the $\Omega_{m}-w_{0}$ confidence contours obtained with the different positional uncertainties for the JBP model using \textit{Ares}  SL data. This Figure clearly shows that increasing the positional uncertainty is translated in an enlargement of the confidence contours and a systematic shift in the $\Omega_{m0}$ estimation towards the fiducial value. 

However, the $\chi^{2}$ statistical estimator also depends on the uncertainty considered for the position of the multiple images.
Thus, the change in the shape of the confidence contours at $0.25\arcsec$ could be explained by the underestimation of the image position uncertainties. For instance, \cite{Acebron2017}  measure an average positional accuracy of $0.66\arcsec$ for Ares using a strong lens model under the standard cosmology. 
Therefore, in the case of our $w(z)$ parametrizations, we would expect reasonable Ares models when the image position errors are roughly similar to the average positional accuracy obtained  (i.e. $\sim0.5\arcsec-1\arcsec$).
We confirm that more statistically significant constraints for the lens (cluster) model are obtained for larger errors, i.e., the reduced $\chi^{2}_{red}$ value trending to one.
This same trend is recovered for the JBP, FSLL I, FSLL II and SeLa parameterizations (see Tables \ref{tab:BA}-\ref{tab:SeLa}). 

The main indicator for the quality of the cosmological constraints is provided by the FOM values. Although there is not a clear tendency FOM vs. $\delta_{pos}$, the strong constraints for the BA, FSLL I, FSLL II and SeLa parameters are obtained when $\delta_{pos}=0.25\arcsec$. However, this uncertainty can lead to a poor cluster model. Moreover, this error provide SL confidence contours only consistent within the $3\sigma$ confidence levels with those obtained from the other cosmological tests. The optimum fit is obtained as a compromise  between those $\chi_{red}$ and RMS values that provide a good lens model and the FOM value that gives cosmological constraints that are also in agreement with other probes. Thus, in the following, we discuss the parameter estimation for the case in which those criteria are fulfilled (i.e. $\delta_{pos}=1\arcsec$, see Figure \ref{fig:JBPmodelsl10} and Appendix \ref{sec:AppB}).

\subsection{$w(z)$ parameter estimations from SL in Abell 1689}
\begin{figure}
\figurenum{2}
\plotone{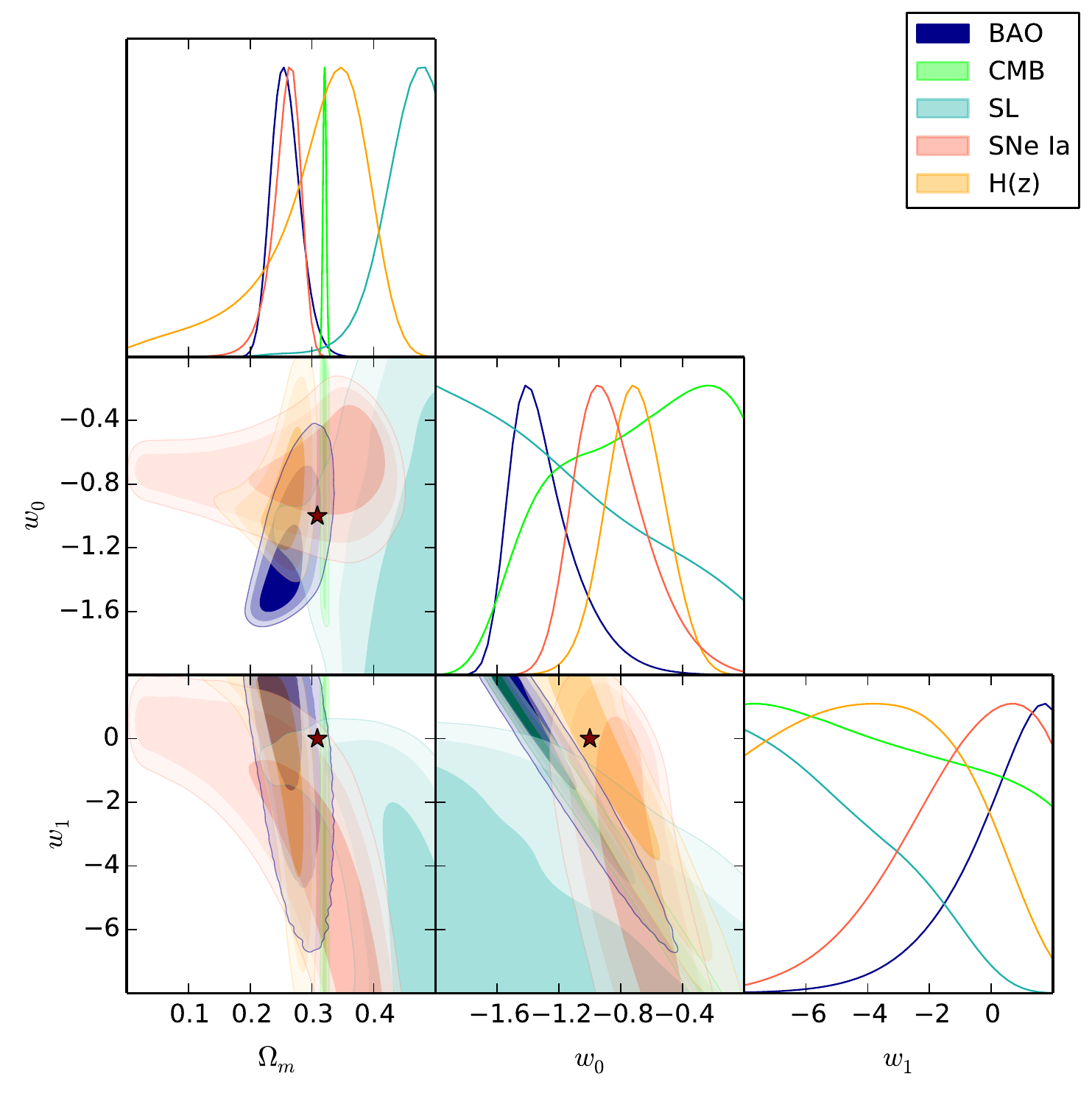}
\caption{Confidence contours (1$\sigma$, 2$\sigma$, 3$\sigma$) and the marginalized 1-dimensional posterior probability distributions on the $\Omega_{m}$, $w_{0}$ and $w_{1}$ parameters for the
cosmological model JBP for Abell 1689 with $\delta_{pos}=0.25\arcsec$. The star indicates the cosmological parameters as constrained by \cite{Planck:2015XIV} for a $\mathrm{\Lambda_{CDM}}$ cosmology.}
\label{fig:JBPmodelsl025}
\end{figure}

\begin{figure}
\figurenum{3}
\plotone{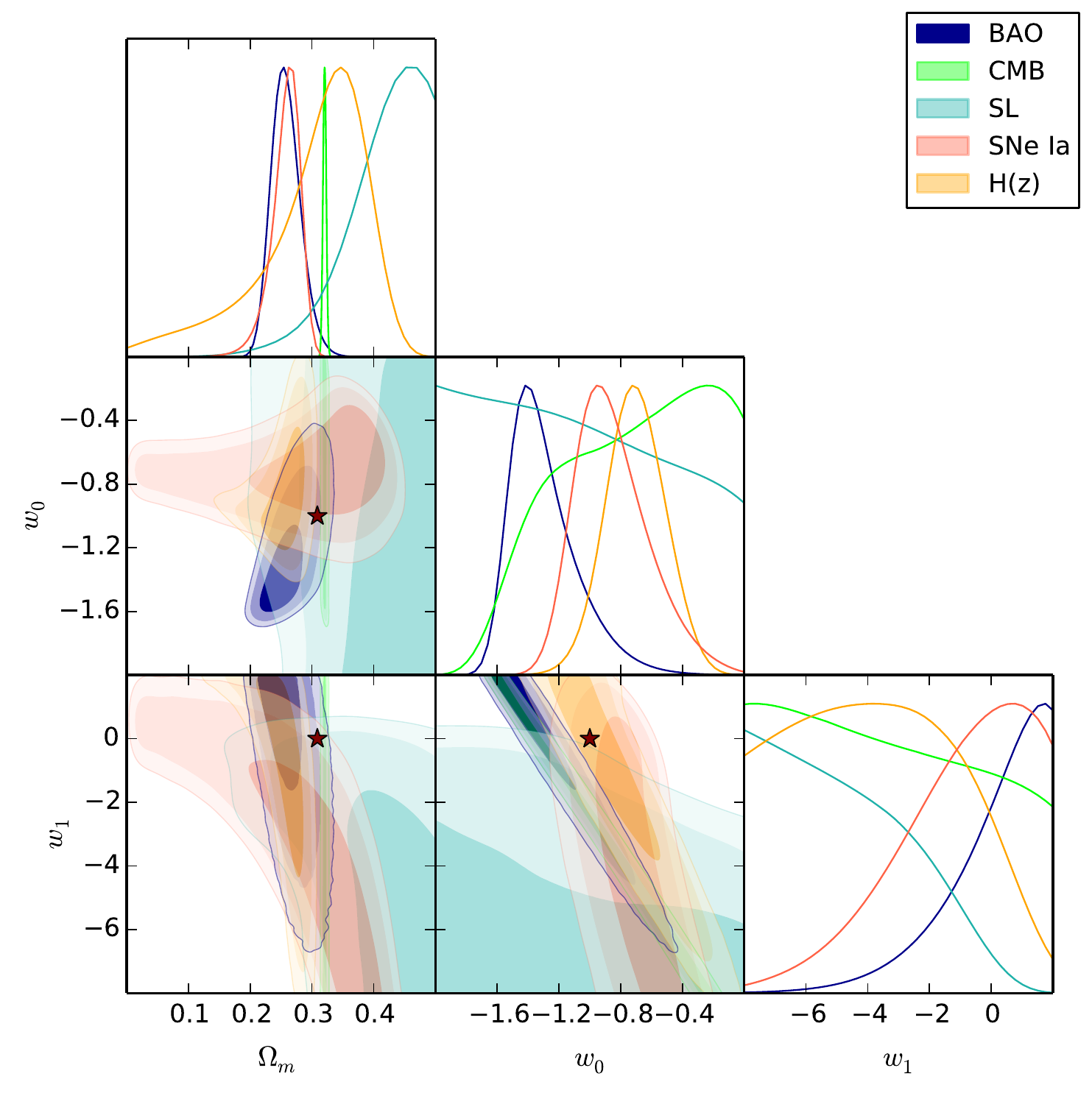}
\caption{Confidence contours (1$\sigma$, 2$\sigma$, 3$\sigma$) and the marginalized 1-dimensional posterior probability distributions on the $\Omega_{m}$, $w_{0}$ and $w_{1}$ parameters for the
cosmological model JBP for Abell 1689 with $\delta_{pos}=0.5\arcsec$. The star indicates the cosmological parameters as constrained by \cite{Planck:2015XIV} for a $\mathrm{\Lambda_{CDM}}$ cosmology.}
\label{fig:JBPmodelsl05}
\end{figure}

\begin{figure}
\figurenum{4}
\plotone{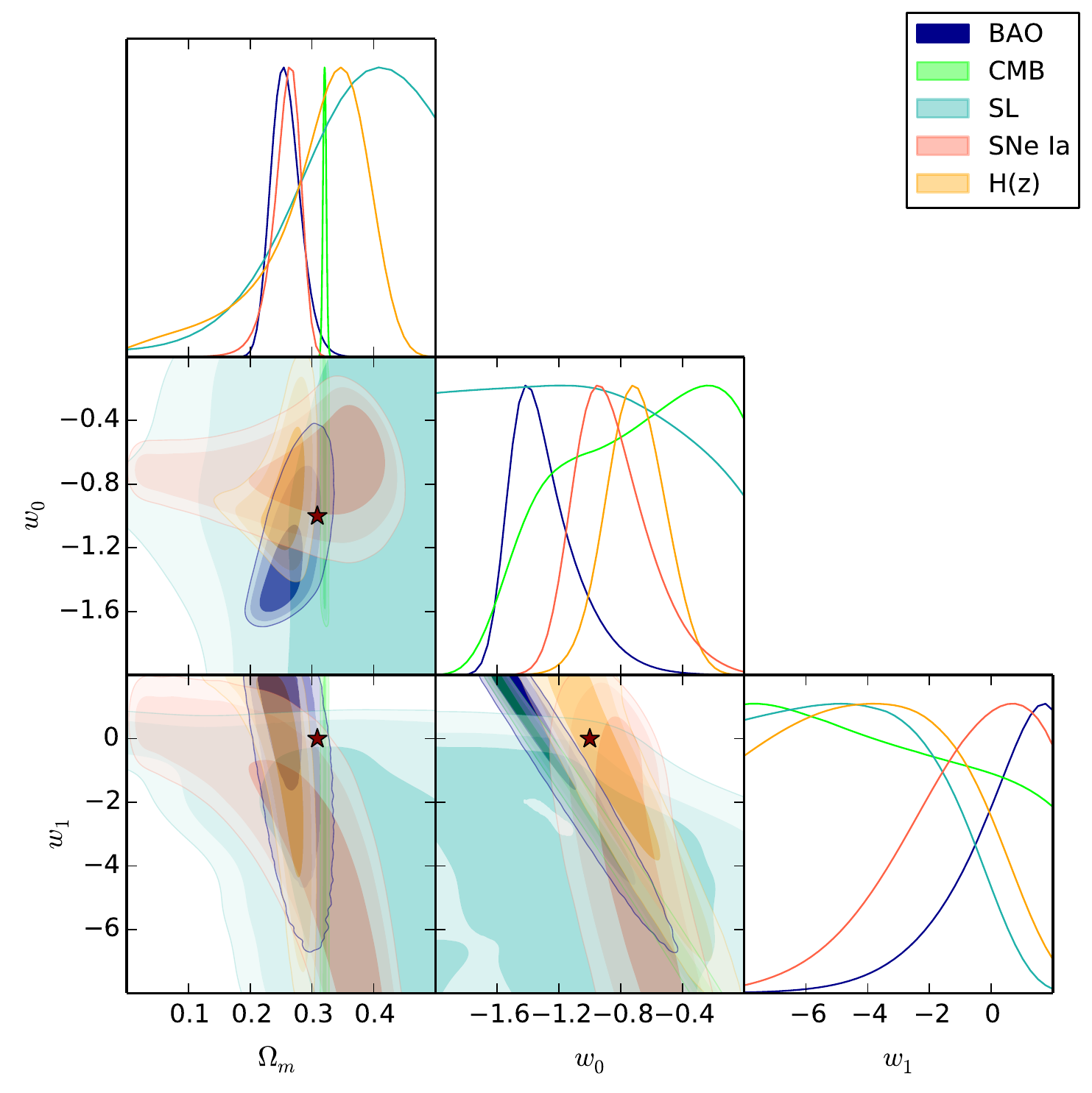}
\caption{Confidence contours (1$\sigma$, 2$\sigma$, 3$\sigma$) and the marginalized 1-dimensional posterior probability distributions on the $\Omega_{m}$, $w_{0}$ and $w_{1}$ parameters for the
cosmological model JBP for Abell 1689 with $\delta_{pos}=1.0\arcsec$. The star indicates the cosmological parameters as constrained by \cite{Planck:2015XIV} for a $\mathrm{\Lambda_{CDM}}$ cosmology.}
\label{fig:JBPmodelsl10}
\end{figure}

In general, for all models, we found that the SL technique using \textit{Abell 1689} data provides better $\Omega_{m}$ constraints than the ones on the equation of state parameters and confirm our previous result: a larger error ($\delta_{pos}=1\arcsec$) provides more significant constraints for the cluster parameters, i.e. $\chi_{red}\sim 1$, and reasonable RMS values. As in the Ares case, the right panel of the Figure \ref{fig:JBPmodelcompare} shows that increasing $\delta_{pos}$ is translated into an enlargement of the confidence contours and a systematic shift in the $\Omega_{m0}$ estimation towards the fiducial value. In addition, although this uncertainty produce the lowest FOM values (i.e. less significant cosmological parameters) for all $w(z)$ parametrizations, the confidence contours are in complete agreement with those of the other probes. The Figures \ref{fig:JBPmodelsl025}-\ref{fig:JBPmodelsl10} show the 1$\sigma$, 2$\sigma$, 3$\sigma$ confidence contours and the marginalized 1-dimensional posterior probability distributions on the $\Omega_m$, $w_0$, $w_1$ parameters for the cosmological model JBP using \textit{Abell 1689} SL data for each positional uncertainty considered. We note again that, when the error in the image position is increased, the $\Omega_{m}-w_{0}$ and $\Omega_{m}-w_{1}$ (or $w_{0}-w_{1}$) confidence contours shift towards the left (upper) region, where the confidence contours from BAO, CMB, SNe Ia, and H(z) probes are overlapped. This same trend is recovered for the BA, FSLL I, FSLL II and SeLa parameterizations (their confidence contours are provided in the Figs. \ref{fig:BAmodelsl10}-\ref{fig:SeLamodelsl10} of the Appendix \ref{sec:AppB}).  

On the other hand, the $w_{0}$ and $w_{1}$ mean values for the
five $w(z)$ parameterizations could suggest a dynamical
equation of state, which can be associated to thawing or freezing quintessence DE \citep{Pantazis:2016}. Nevertheless, all our EoS constraints are consistent with the cosmological constant, i.e. $w_{0}=-1$, and $w_{1}=0$, within the $3\sigma$ confidence level. In addition, there is no significant difference among the $\chi^2_{red}$ and RMS values for different $w(z)$ parameterizations. Therefore, any parametric DE model could be the source of the late cosmic acceleration. We confirm this result in the left panel of the Figure \ref{fig:qz_models_a1689} which shows the reconstruction of the cosmological evolution for each parameterization using the mean values obtained from the SL modeling in Abell 1689 when $\delta_{pos}=1\arcsec$ is considered.

\subsection{Deceleration parameter}
\begin{figure*}
\figurenum{5}
\gridline{\fig{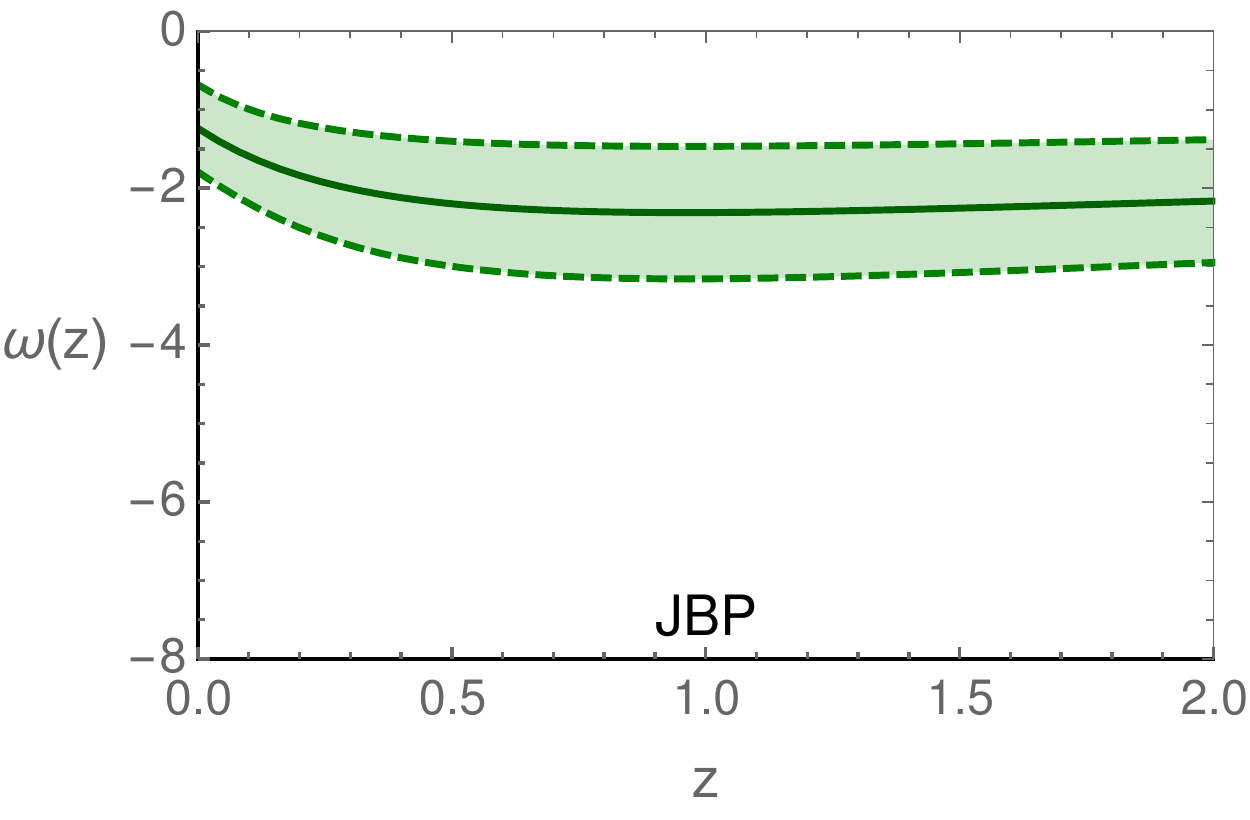}{0.33\textwidth}{}
          \fig{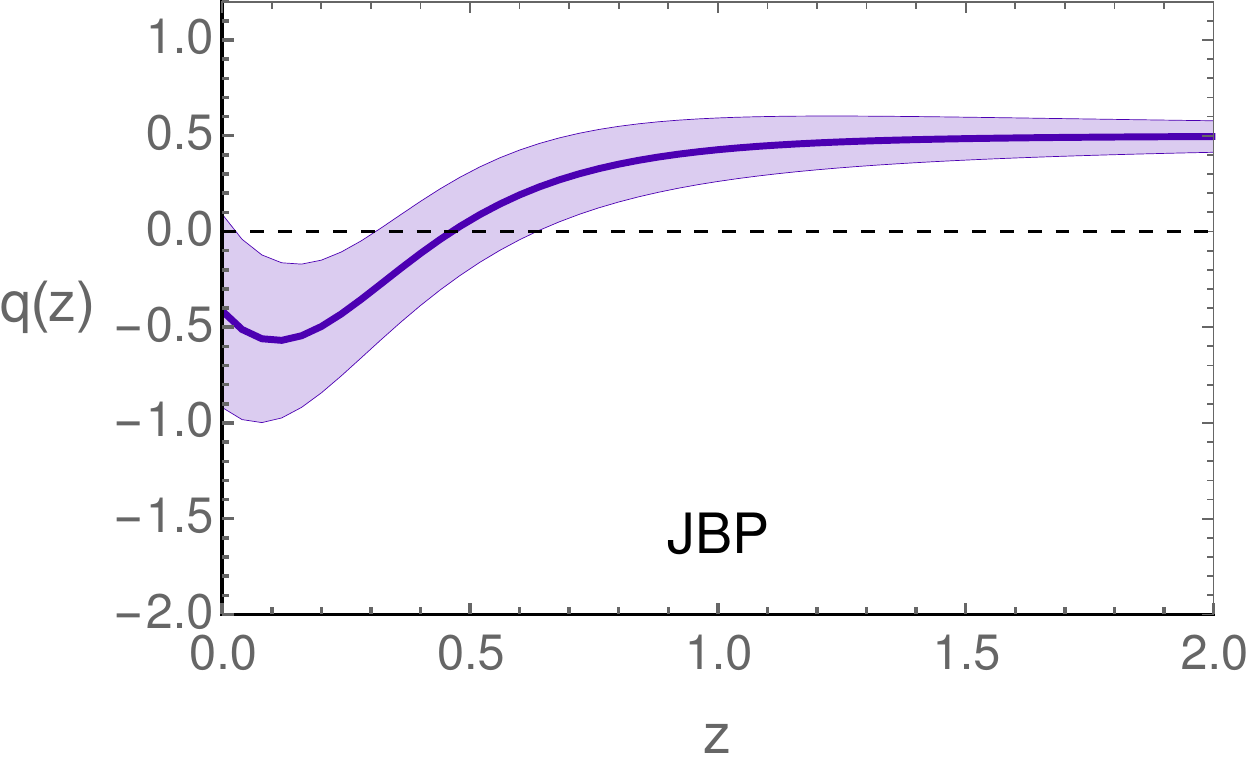}{0.33\textwidth}{}
          }
\gridline{\fig{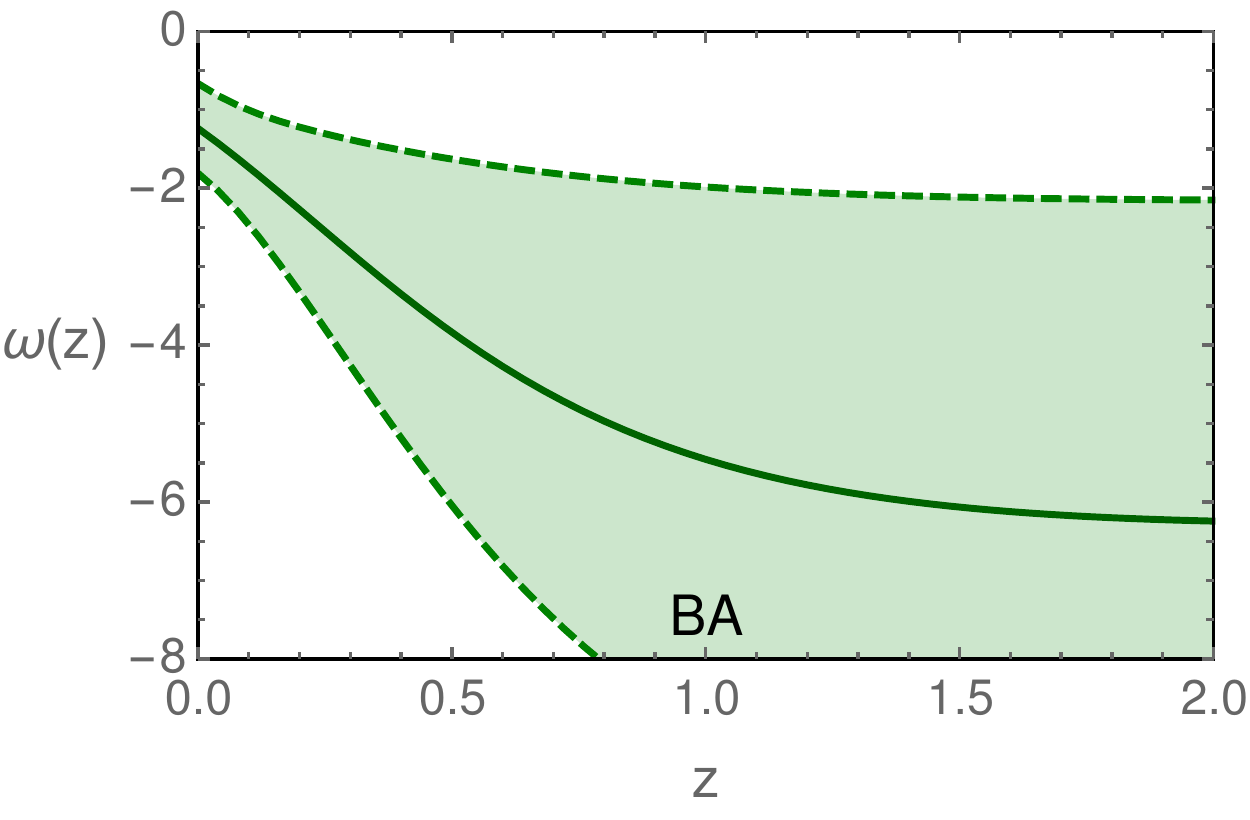}{0.33\textwidth}{}
         \fig{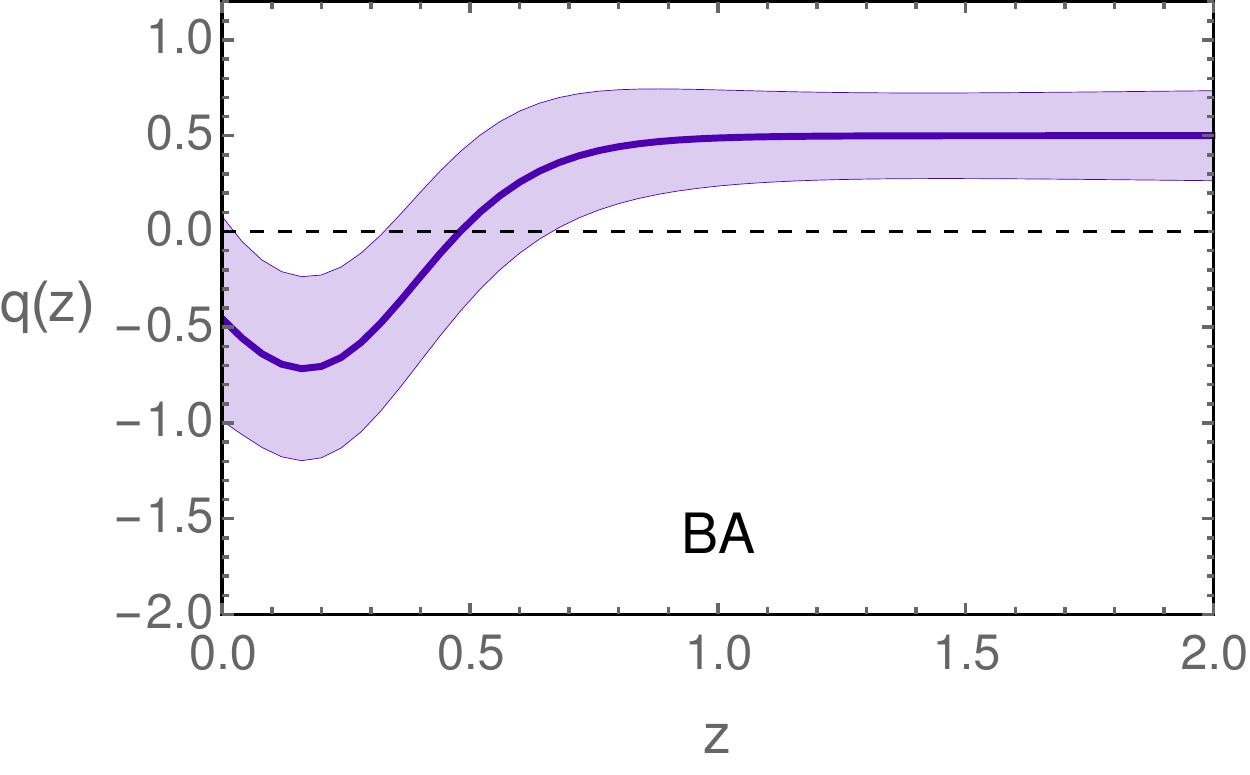}{0.33\textwidth}{}
          }
\gridline{\fig{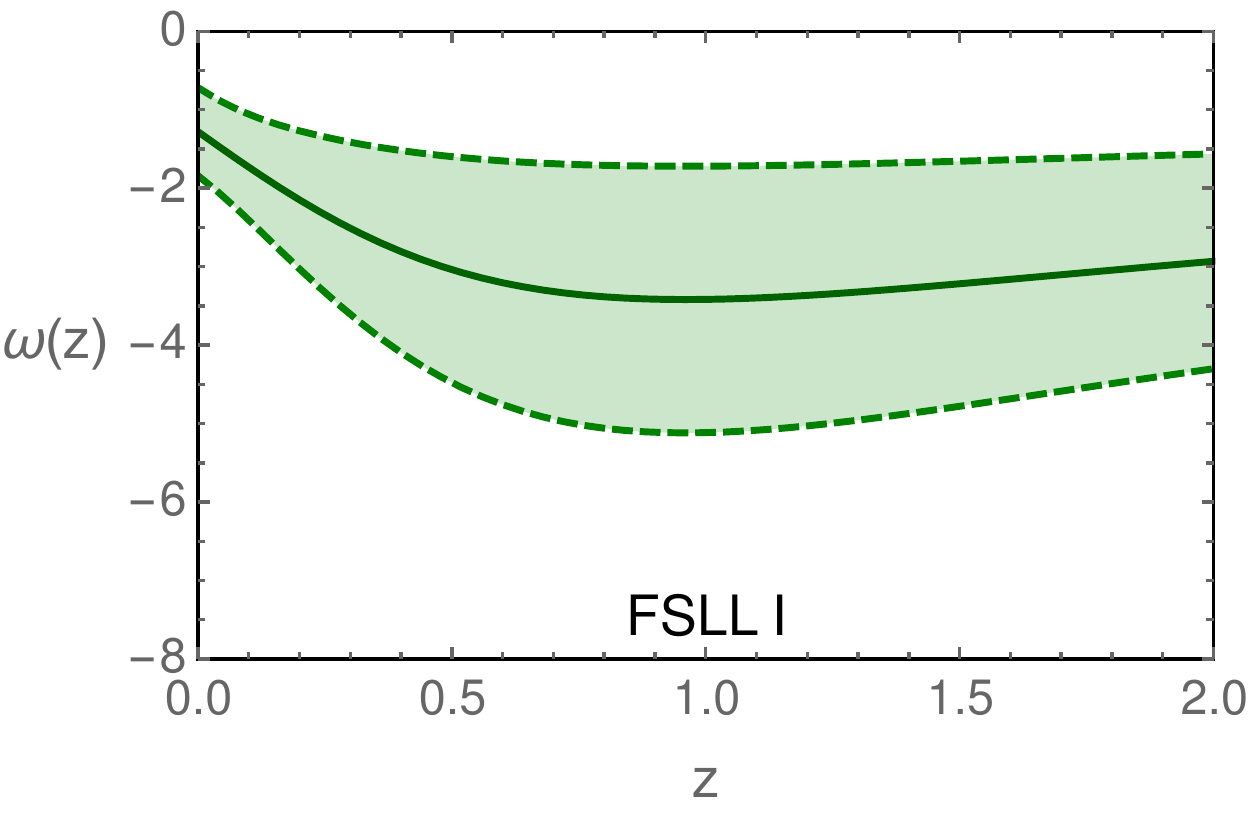}{0.33\textwidth}{}
          \fig{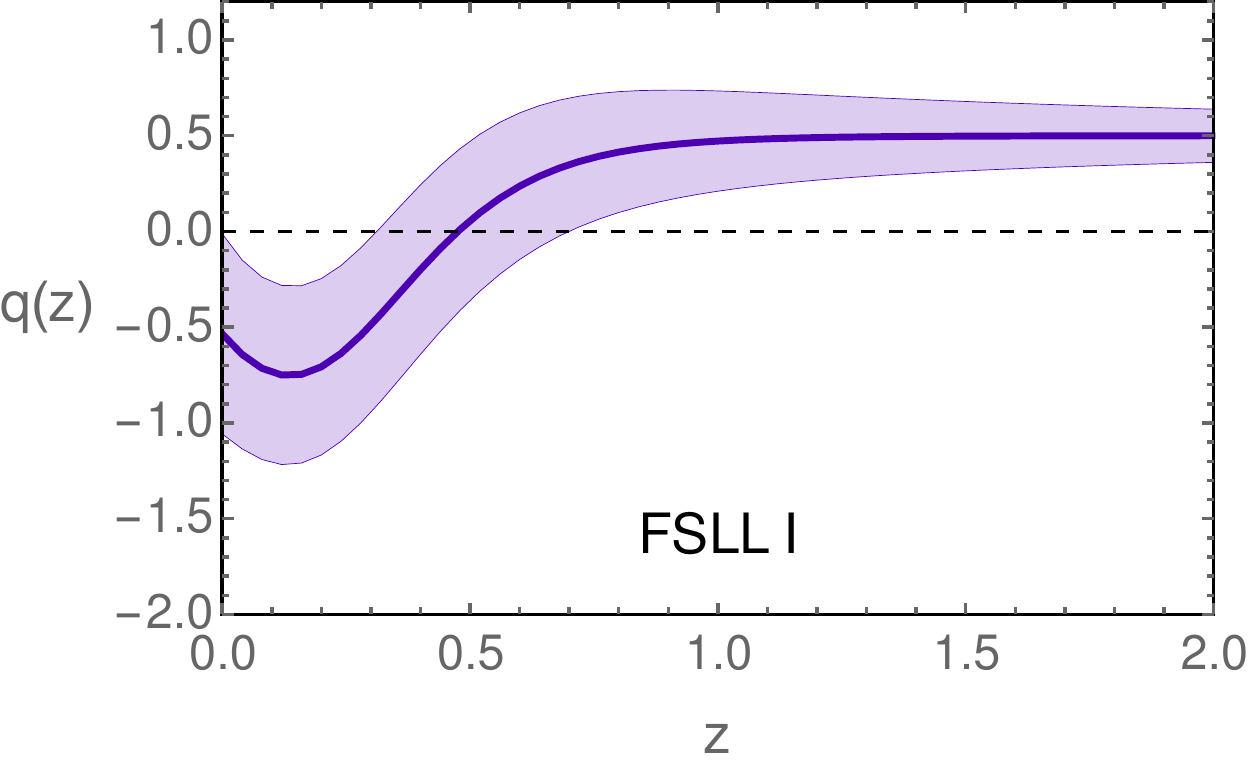}{0.33\textwidth}{}
          }
\gridline{\fig{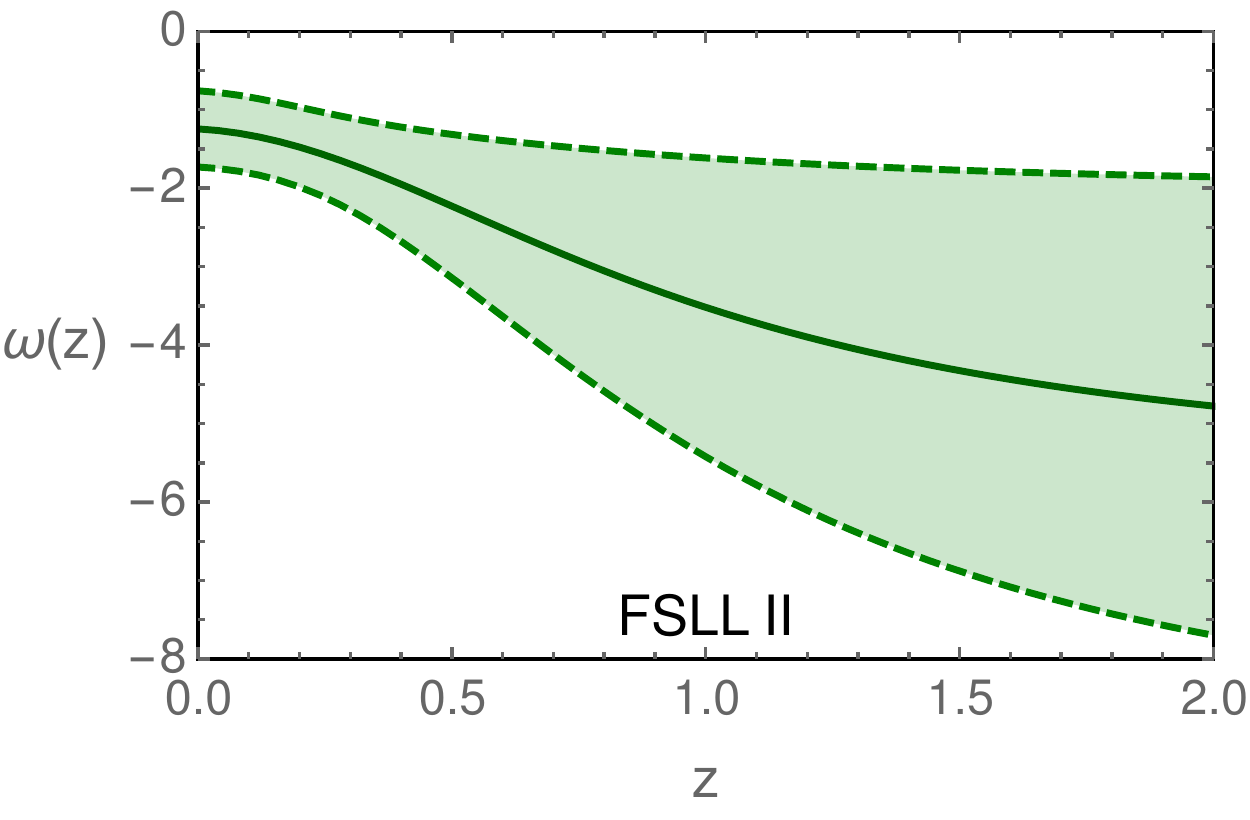}{0.33\textwidth}{}
          \fig{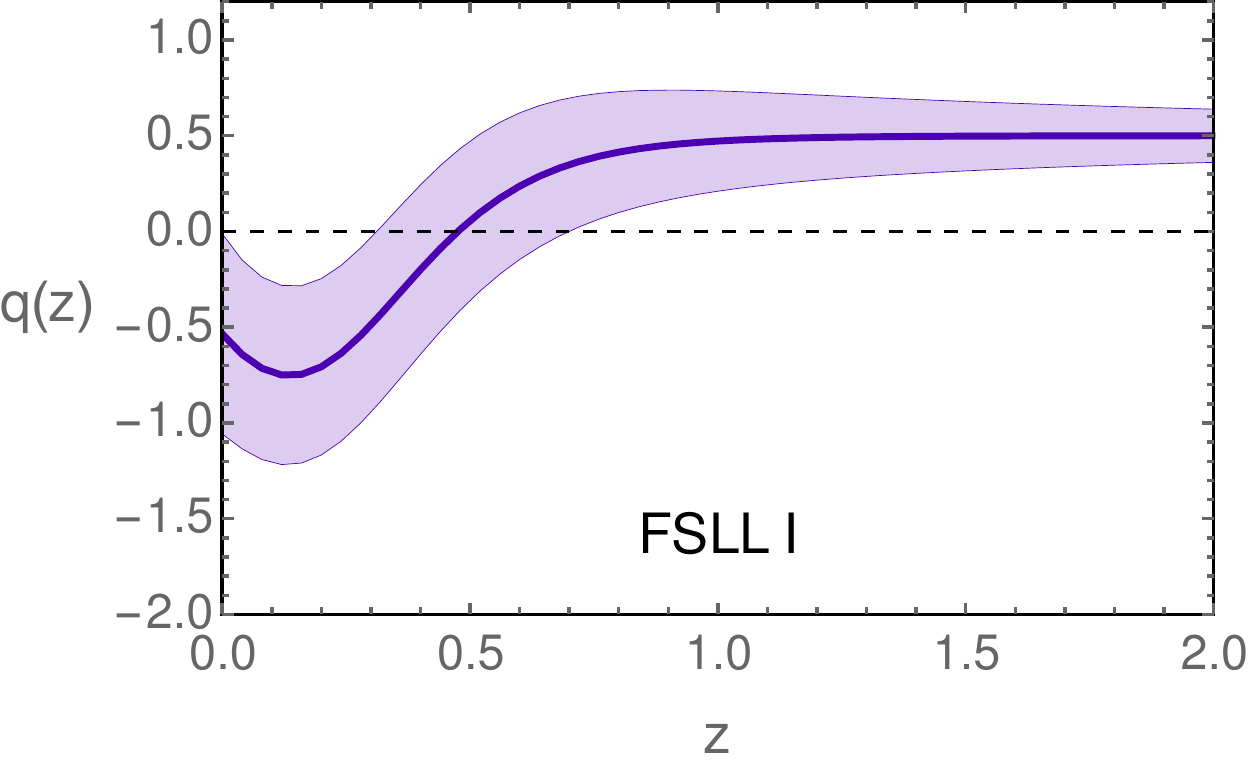}{0.33\textwidth}{}
          }
\gridline{\fig{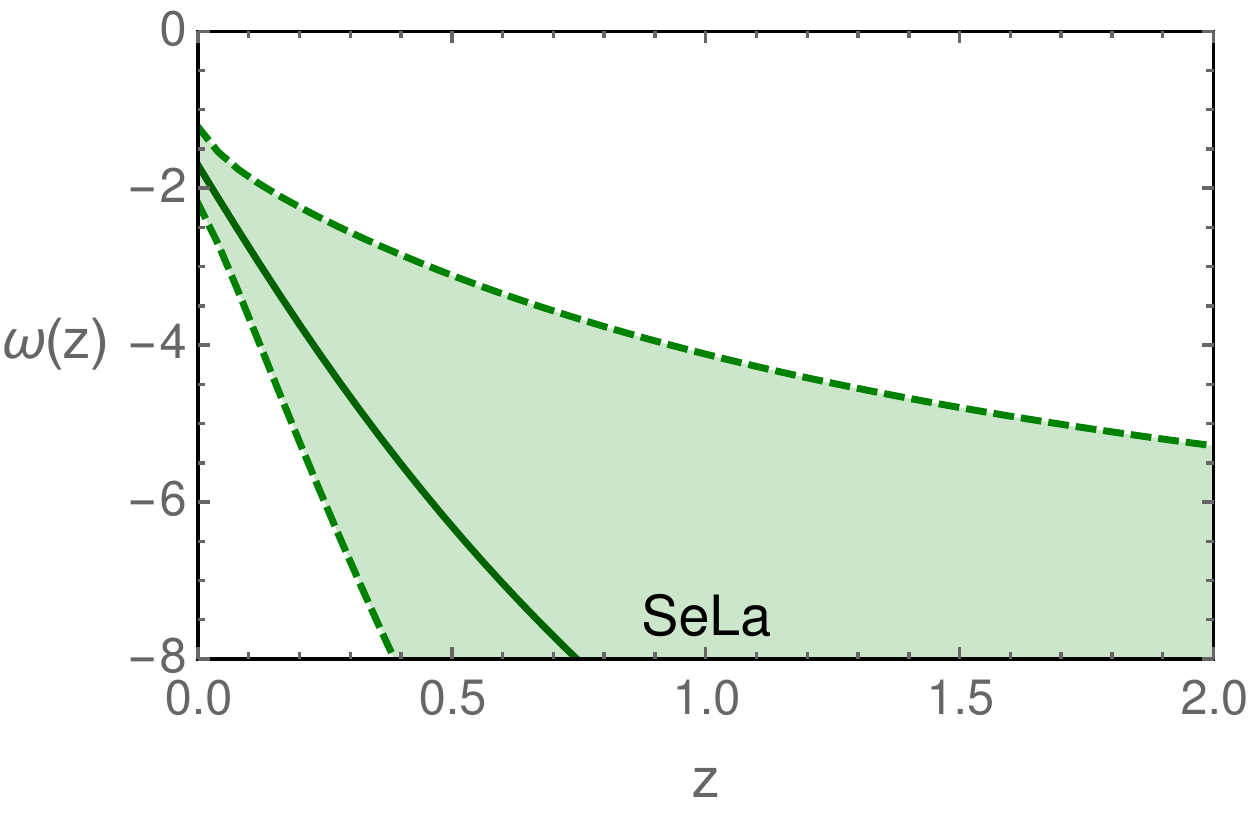}{0.33\textwidth}{}
          \fig{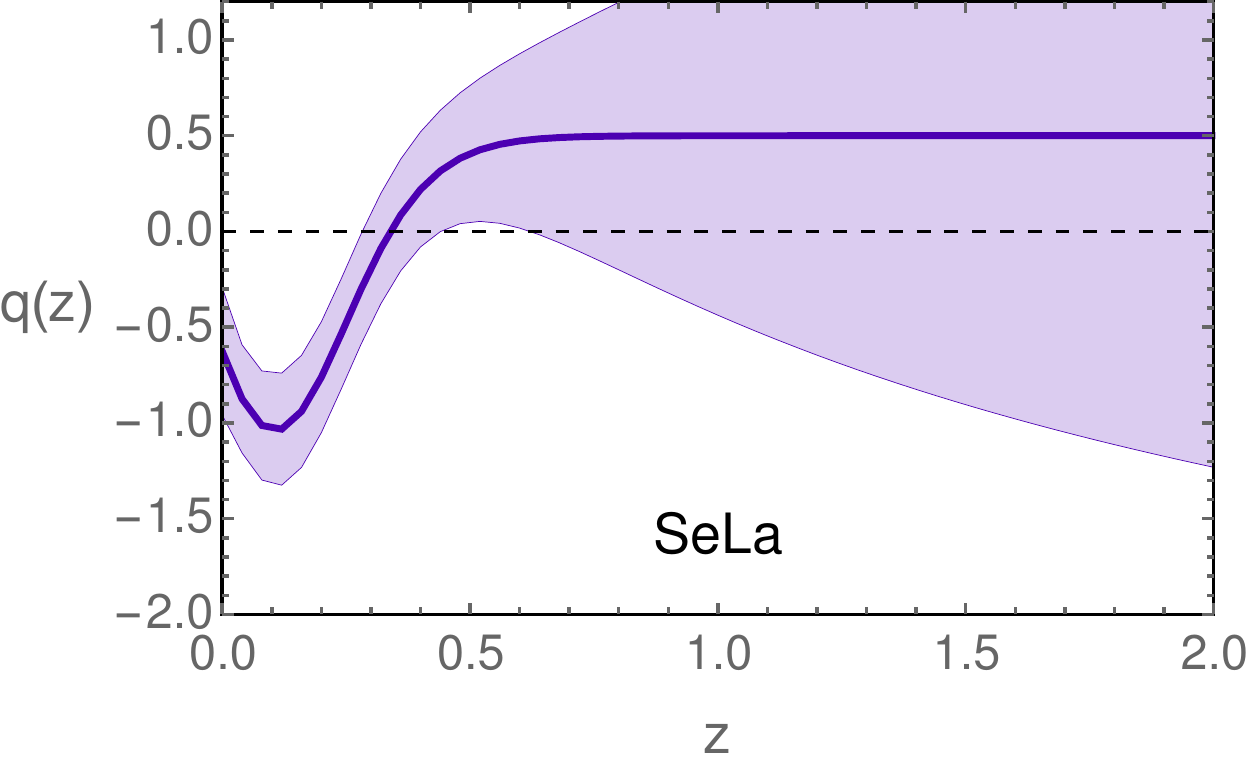}{0.33\textwidth}{}
          }
\caption{Reconstructed equation of state (left panel) and deceleration parameter (right panel) vs. redshift for each $w(z)$ 
parameterization using the mean values obtained from SL data in \textit{Abell 1689} with an image-position error of $1.0\arcsec$.
The shadow regions show the $1\sigma$ region calculated with a MCMC error propagation approach using the SL posterior constraints.}  
\label{fig:qz_models_a1689}
\end{figure*}

The cosmological behavior of the deceleration parameter (Eq. \ref{eq:qz})
is an important test to know whether a DE model is able to handle the
late cosmic acceleration. The right panels of Figure \ref{fig:qz_models_a1689} shows the reconstructed $q(z)$ evolution for each parameterization obtained from \textit{Abell 1689} SL data when the multiple image-positional error is $1\arcsec$. We also have propagated its error within the $1\sigma$ confidence level using a Monte Carlo approach. Notice that the five cosmological models predict an accelerating expansion at late times. 
The transition redshifts, i.e. when the Universe passes from an decelerated phase to one accelerated, are 
$z_{t}={0.44^{+0.21}_{-0.17},0.44^{+0.17}_{-0.12},0.45^{+0.18}_{-0.13},0.50^{+0.22}_{-0.24},0.34^{+0.07}_{-0.04}}$  for the JBP, BA, FSLL I, FSLL II, and SeLa parameterizations, respectively. Furthermore, the $q(z)$ shape for each parameterization is consistent with that of the cosmological constant within the $1\sigma$ confidence level.
\floattable
\tablenum{1}
\label{tabJBP}
\begin{deluxetable}{cccccccc}
\tablecaption{JBP mean fit parameters obtained for both galaxy clusters SL data. The columns give the reduced $\chi^{2}_{red}$, RMS in the image plane (in arcseconds) and the mean values for $\Omega_{m}$, $w_{0}$, $w_{1}$ with $68\%$ confidence level errors. The mean values estimated with H(z), SNIa, BAO and CMB data are also provided.}
\tablecolumns{8}
\tablewidth{0pt}
\tablehead{
\colhead{Cluster name} &
\colhead{Error in the pos. ($\arcsec$)} &
\colhead{$\chi^{2}_{red}$} &
\colhead{RMS ($\arcsec$)} &
\colhead{FOM} &
\colhead{$\Omega_{m}$} &
\colhead{$w_{0}$} &
\colhead{$w_{1}$} 
}
\startdata
Abell\,1689 & 0.25 & 11.37 & 0.54 & 15.09 & $0.47_{-0.06}^{+0.04}$ &  $-1.29_{-0.01}^{+1.04}$& $-6.46_{-0.33}^{+4.88}$\\  
\textit{Ares} & &2.73& 0.59& 127.92& $0.37_{-0.11}^{+0.12}$&$-0.82_{-0.11}^{+0.11}$&$-0.14_{-0.04}^{+0.36}$
	\vspace{0.2cm}\\
Abell\,1689 & 0.5  & 3.14 & 0.64& 8.13 &$0.46_{-0.08}^{+0.08}$&  $-1.11 _{-0.04}^{+1.20}$ & $-6.14_{-0.07}^{+5.49}$\\
\textit{Ares} & &0.78 &0.65&11.83  &$0.46_{-0.11}^{+0.06}$&$-0.90_{-0.49}^{+0.60}$&$-1.77_{-3.01}^{+0.07}$
	\vspace{0.2cm}\\
Abell\,1689 & 1.0  & 0.95& 0.88& 4.15& $0.43_{-0.12}^{+0.18}$& $-1.07_{-0.60}^{+0.69}$& $-5.09_{-3.40}^{+3.07}$\\
\textit{Ares} &&0.77& 0.94& 20.46 &$0.33_{-0.06}^{+0.06}$ & $-1.06_{-0.43}^{+0.83}$& $-5.81_{-3.16}^{+3.41}$\\
\multicolumn{8}{c}{Complementary probes} \\
H(z) & ---& $0.54$ & --- & 354.12 &$0.26^{+0.01}_{-0.02}$ & $-0.88^{+0.27}_{-0.19}$ & $-0.70^{+1.83}_{-2.59}$\\
SNIa & ---& $0.98$ & --- & 75.51& $0.32^{+0.05}_{-0.10}$ & $-0.70^{+0.19}_{-0.18}$ & $-4.44^{+3.40}_{-3.49}$\\
BAO & ---& $2.17$ & --- & 646.43 &$0.25^{+0.02}_{-0.02}$ & $-1.34^{+0.26}_{-0.15}$ & $0.43^{+1.13}_{-1.99}$\\
CMB & ---& $58.8$ & --- & 8207.46 &$0.32^{+0.002}_{-0.002}$ & $-0.69^{+0.50}_{-0.68}$ & $-4.54^{+4.26}_{-3.66}$\\
\enddata
\end{deluxetable}
\floattable
\tablenum{2}
\label{tab:BA}
\begin{deluxetable}{cccccccc}
\tablecaption{The same as Table \ref{tabJBP} for the BA parameterization.}
\tablecolumns{8}
\tablewidth{0pt}
\tablehead{
\colhead{Cluster name} &
\colhead{Error in the pos. ($\arcsec$)} &
\colhead{$\chi^{2}_{red}$} &
\colhead{RMS ($\arcsec$)} &
\colhead{FOM} &
\colhead{$\Omega_{m}$} &
\colhead{$w_{0}$} &
\colhead{$w_{1}$} 
}
\startdata
Abell\,1689 & 0.25" & 11.40 & 0.54& 13.57&$0.47_{-0.06}^{+0.05}$ &  $-1.32_{-0.01}^{+0.98}$& $-6.28_{-0.07}^{+5.31}$\\ 
\textit{Ares} & &2.69& 0.59& 5.25& $0.27_{-0.01}^{+0.46}$ & $-0.89_{-0.48}^{+0.16}$ & $0.09_{-1.10}^{+0.91}$
	\vspace{0.2cm}\\
Abell\,1689 & 0.5"  & 3.15 & 0.64 & 6.22& $0.44_{-0.10}^{+0.08}$ &  $-1.15_{-0.02}^{+1.20}$& $-5.22_{-0.57}^{+6.17}$\\
\textit{Ares} & &0.80& 0.65 & 8.25& $0.20_{-0.01}^{+0.36}$ & $-1.10_{-0.24}^{+0.31}$ & $0.36_{-1.81}^{+0.16}$
	\vspace{0.2cm}\\
Abell\,1689 & 1.0"  & 0.94 &  0.89& 3.26&$0.41_{-0.15}^{+0.17}$ &  $-1.08_{-0.02}^{+1.26}$& $-4.49_{-7.69}^{+0.62}$\\
\textit{Ares} & &0.33& 0.93& 6.91&$0.26_{-0.0}^{+0.31}$ & $-1.37_{-0.24}^{+0.31}$ & $-1.88_{-6.28}^{+0.18}$\\
\multicolumn{8}{c}{Complementary probes} \\
H(z) & ---& $0.57$ & --- & 405.21 &$0.25^{+0.02}_{-0.08}$ & $-0.90^{+0.13}_{-0.12}$ & $0.01^{+0.43}_{-0.85}$\\
SNIa & ---& $0.98$ & --- & 70.47&$0.37^{+0.04}_{-0.09}$ & $-0.78^{+0.22}_{-0.17}$ & $-3.24^{+2.37}_{-3.14}$\\
BAO & ---& $2.25$ & --- & 1541.69&$0.26^{+0.02}_{-0.02}$ & $-1.23^{+0.18}_{-0.16}$ & $-0.23^{+0.52}_{-0.69}$\\
CMB & ---& $58.8$ & --- &3492.21 &$0.32^{+0.002}_{-0.002}$ & $-0.63^{+0.45}_{-0.67}$ & $-2.24^{+2.18}_{-1.82}$\\
\enddata
\end{deluxetable}
\floattable
\tablenum{3}
\label{tab:FSLLI}
\begin{deluxetable}{cccccCcc}
\tablecaption{The same as Table \ref{tabJBP} for the FSLL I parameterization.}
\tablecolumns{8}
\tablewidth{0pt}
\tablehead{
\colhead{Cluster name} &
\colhead{Error in the pos. ($\arcsec$)} &
\colhead{$\chi^{2}_{red}$} &
\colhead{RMS ($\arcsec$)} &
\colhead{FOM} &
\colhead{$\Omega_{m}$} &
\colhead{$w_{0}$} &
\colhead{$w_{1}$} 
}
\startdata
Abell\,1689 & 0.25" &11.53&0.54&12.67 &$0.46_{-0.06}^{+0.05}$ &  $-1.39_{-0.04}^{+0.93}$& $-6.60_{-0.07}^{+5.04}$\\ 
	\textit{Ares} & &2.70&0.59&168.03&$0.17_{-0.03}^{+0.22}$&$-0.99_{-0.16}^{+0.24}$ &$0.48_{-0.72}^{+0.29}$
    \vspace{0.2cm}\\
	Abell\,1689 & 0.5"  & 3.07&0.64& 4.10&$0.42_{-0.12}^{+0.09}$ &  $-1.17_{-0.21}^{+1.08}$& $-4.75_{-8.73}^{+0.14}$\\
	\textit{Ares} & &0.80&0.65&11.29 &$0.23_{-0.02}^{+0.36}$&$-1.36_{-0.15}^{+0.60}$&$0.79_{-3.00}^{+0.08}$
    \vspace{0.2cm}\\
	Abell\,1689 & 1.0"  & 0.93&0.89& 3.34& $0.40_{-0.15}^{+0.17}$ &  $-1.15_{-0.16}^{+1.07}$& $-4.99_{-1.12}^{+6.08}$\\
	\textit{Ares} & &0.33&0.92& 8.03&$0.27_{-0.01}^{+0.32}$&$-1.45_{-0.15}^{+0.97}$&$-1.89_{-7.13}^{+0.13}$\\
    \multicolumn{8}{c}{Complementary probes} \\
H(z) & ---& $0.48$ & --- & 268.56 &$0.25^{+0.02}_{-0.06}$ & $-0.95^{+0.19}_{-0.15}$ & $0.14^{+1.03}_{-1.46}$\\
SNIa & ---& $0.98$ & --- &71.57 &$0.36^{+0.05}_{-0.10}$ & $-0.74^{+0.22}_{-0.18}$ & $-3.77^{+2.77}_{-3.37}$\\
BAO & ---& $2.13$ & --- & 1014.38&$0.24^{+0.02}_{-0.02}$ & $-1.40^{+0.25}_{-0.19}$ & $0.51^{+0.92}_{-1.24}$\\
CMB & ---& $58.8$ & --- & 5455.18&$0.32^{+0.002}_{-0.002}$ & $-0.69^{+0.50}_{-0.68}$ & $-2.86^{+3.05}_{-2.54}$\\
\enddata
\end{deluxetable}
\floattable
\tablenum{4}
\label{tab:FSLLII}
\begin{deluxetable}{cccccccc}
\tablecaption{The same as Table \ref{tabJBP} for the FSLL II parameterization.}
\tablecolumns{8}
\tablewidth{0pt}
\tablehead{
\colhead{Cluster name} &
\colhead{Error in the pos. ($\arcsec$)} &
\colhead{$\chi^{2}_{red}$} &
\colhead{RMS ($\arcsec$)} &
\colhead{FOM} &
\colhead{$\Omega_{m}$} &
\colhead{$w_{0}$} &
\colhead{$w_{1}$} 
}
\startdata
Abell\,1689 & 0.25" & 11.90& 0.54& 19.78&$0.41_{-0.06}^{+0.05}$ &  $-1.61_{-0.03}^{+0.54}$& $-5.59_{-0.09}^{+6.82}$\\ 
\textit{Ares} & &2.71&0.59& 50.04&$0.19_{-0.01}^{+0.33}$&$-0.81_{-0.10}^{+0.06}$&$0.18_{-0.80}^{+0.13}$
	\vspace{0.2cm}\\
Abell\,1689 & 0.5"  & 3.17&0.64& 7.96&$0.38_{-0.10}^{+0.09}$ &  $-1.45_{-0.26}^{+0.57}$& $-4.81_{-0.29}^{+7.24}$\\
\textit{Ares} & &0.79&0.65& 8.74&$0.38_{-0.21}^{+0.15}$&$-1.00_{-0.44}^{+0.25}$&$-0.64_{-4.69}^{+0.38}$
	\vspace{0.2cm}\\
Abell\,1689 & 1.0"  &  0.94&0.89& 3.31&$0.35_{-0.14}^{+0.20}$ &  $-1.29_{-0.31}^{+0.70}$& $-4.78_{-0.10}^{+7.57}$\\
\textit{Ares} & &0.32&0.92& 9.50&$0.28_{-0.14}^{+0.06}$&$-1.35_{-0.54}^{+0.29}$&$-2.11_{-7.14}^{+0.50}$\\
\multicolumn{8}{c}{Complementary probes} \\
H(z) & --- & $0.56$ & --- & 164.36 & $0.26^{+0.02}_{-0.06}$ & $-0.92^{+0.11}_{-0.10}$ & $-0.30^{+0.92}_{-2.12}$\\
SNIa & --- & $0.98$ & --- & 71.30 &$0.32^{+0.04}_{-0.11}$ & $-0.98^{+0.15}_{-0.15}$ & $-4.34^{+3.90}_{-3.86}$\\
BAO & --- & $2.21$ & --- & 530.10 &$0.27^{+0.02}_{-0.02}$ & $-1.19^{+0.12}_{-0.10}$ & $-1.34^{+1.23}_{-1.91}$\\
CMB & --- & $58.8$ & --- &2702.46 &$0.32^{+0.002}_{-0.002}$ & $-0.87^{+0.22}_{-0.34}$ & $-4.96^{+4.01}_{-3.51}$\\
\enddata
\end{deluxetable}

\floattable
\tablenum{6}
\label{tab:SeLa}
\begin{deluxetable}{cccccccc}
\tablecaption{The same as Table \ref{tabJBP} for the SeLa parameterization.}
\tablecolumns{7}
\tablewidth{0pt}
\tablehead{
\colhead{Cluster name} &
\colhead{Error in the pos. ($\arcsec$)} &
\colhead{$\chi^{2}_{red}$} &
\colhead{RMS ($\arcsec$)} &
\colhead{FOM} &
\colhead{$\Omega_{m}$} &
\colhead{$w_{0}$} &
\colhead{$w_{0.5}$} 
}
\startdata
Abell\,1689 & 0.25" &  11.66&  0.54& 43.94 &$0.42_{-0.10}^{+0.07}$ &  $-1.23_{-0.24}^{+0.94}$& $0.61_{-1.57}^{+0.38}$\\ 
\textit{Ares} & &  6.01&  4.80 & 33.84 & $0.80_{-0.09}^{+0.06}$ &  $-1.04_{-0.08}^{+1.25}$& $0.38_{-1.56}^{+0.65}$
\vspace{0.2cm}   \\
Abell\,1689 & 0.5"  & 3.06 & 0.64& 25.21 & $0.41_{-0.09}^{+0.11}$ &  $-1.18_{-0.13}^{+1.06}$& $-0.54_{-1.96}^{+0.02}$\\
\textit{Ares}&  & 1.57 & 1.11 & 25.09 & $0.66_{-0.09}^{+0.12}$ &  $-1.26_{-0.01}^{+1.07}$& $-0.59_{-1.83}^{+0.14}$
\vspace{0.2cm}   \\
Abell\,1689 & 1.0"  &0.93&0.90& 12.67 &$0.46_{-0.05}^{+0.04}$ &  $-1.39_{-0.44}^{+0.73}$& $-6.60_{-2.44}^{+3.87}$\\
\textit{Ares} & &0.48&1.06& 23.61 &$0.41_{-0.10}^{+0.16}$ &  $-1.16_{-0.04}^{+1.17}$& $0.49_{-1.40}^{+0.79}$\\
 \multicolumn{7}{c}{Complementary probes} \\
H(z) & ---& $0.55$ & --- & 703.29&$0.25^{+0.02}_{-0.07}$ & $-0.90^{+0.14}_{-0.12}$ & $-1.01^{+0.31}_{-0.36}$\\
SNIa & ---& $0.98$ & --- & 111.75&$0.37^{+0.05}_{-0.09}$ & $-0.73^{+0.28}_{-0.18}$ & $-2.64^{+1.23}_{-1.57}$\\
BAO & ---& $9.95$ & --- & 377.94&$0.23^{+0.04}_{-0.08}$ & $-1.04^{+0.28}_{-0.23}$ & $-1.11^{+0.55}_{-0.37}$\\
CMB & ---& $58.8$ & --- & 1004.52 &$0.32^{+0.002}_{-0.002}$ & $-0.56^{+0.40}_{-0.54}$ & $-1.98^{+0.51}_{-0.50}$\\
\enddata
\end{deluxetable}

\section{conclusions}\label{sec:conclusions}
Several recent studies have shown that dark energy could deviate from a cosmological constant \citep{Ferreira:2017, Zhao:2017Nat}. A simple way to investigate such alternative dark energy models is to parameterize the dark energy equation of state as a function of redshift. In order to elucidate the nature of dark energy, numerous parameterizations have been proposed \citep[see for instance][and references therein]{Pantazis:2016}. The typical tests to constrain cosmological parameters use SNe Ia, H(z), BAO and CMB distance posterior measurements.
Nevertheless, some of them could provide biased constraints
because either the data or the test fitting formulae are derived assuming an underlying standard cosmology (see Appendix \ref{sec:Appendix}). Furthermore, new complementary techniques could break the degeneracy between parameters and obtain stringent constraints which could help us distinguish the nature of dark energy. 

In this paper, which is the first in a series, we investigate a promising technique to study alternative cosmological models and to constrain their parameters using the strong lensing features in galaxy clusters. This method has the advantage of providing constraints which are not biased due to an underlying cosmology.

We have considered the following five popular bi-parametric CPL-like ansatz: JBP, BA, FSLL I, FSLL II, and SeLa and constrained their parameters using the SL data in a real galaxy cluster, \textit{Abell 1689}, and a simulated one \textit{Ares}. We implemented these $w(z)$ parameterizations in the LENSTOOL code which uses a MCMC algorithm to simultaneously constrain the lens model and the $w(z)$ parameters. In addition, we have considered three different image-positional errors to quantify how the cosmological constraints are affected by these uncertainties in the lens modeling. In general, we found that the SL technique provides competitive constraints on the $w(z)$ parameters in comparison with the common cosmological tests. Moreover, when increasing the image-positional error (from $0.5\arcsec$ to $1.0\arcsec$), we find that systematic biases with respect to the known input cosmological values in the simulated cluster decrease. After taking this calibration into account in the real data, our SL constraints are consistent with those obtained from other probes.

In summary, we have exploited the strong lensing modeling in galaxy clusters as a cosmological probe. Although we have measured competitive constraints on the $w(z)$ parameters, further analysis on the galaxy clusters and their environment is needed to improve the strong lensing modeling and hence to more tightly estimate cosmological parameters. In forthcoming papers, we will test this method to constrain the parameter of other cosmological scenarios, for instance, those considering interactions in the dark sector.

\acknowledgments
We thank the anonymous referee for thoughtful remarks
and suggestions.
J.M. acknowledges the support from CONICYT/FONDECYT project 3160674 and thanks the hospitality of the LAM staff where part of this work was done.
This work has been carried out thanks to the support of the OCEVU Labex (ANR-11-LABX-0060) and the A*MIDEX project (ANR-11-IDEX-0001-02) funded by the "Investissements d'Avenir" French government program managed by the ANR. \\
This work was granted access to the HPC resources of Aix-Marseille Universit\'{e} financed by the project Equip@Meso (ANR-10-EQPX-29-01) of the program "Investissements d'Avenir" supervised by the Agence Nationale pour la Recherche. This research has been carried out thanks to PROGRAMA UNAM-DGAPA-PAPIIT IA102517. T.~V. thanks the staff of the Instituto de F\'{\i}sica y Astronom\'{\i}a of the Universidad de Valpara\'{\i}so. M.L. acknowledges the support from Centre national de la recherche scientifique (CNRS), Programme National de Cosmologie et Galaxies (PNCG) and CNES.

\appendix
\section{Additional cosmological data}
\label{sec:Appendix}
We compare the constraints obtained from the strong lensing modeling with those from BAO, CMB, SNe Ia  and H(z) cosmological probes. In the following we describe briefly these cosmological data, for further details on how their figure-of-merit is constructed see \citet{Magana:2015,Magana:2017} and references therein.

\subsection{BAO}
Large-scale galaxy surveys offer the possibility of measuring the signature of Baryon Acoustic Oscillations which is a typical length scale imprinted in both photons and baryons by the propagation of sound waves in the primordial plasma of the Universe. This signal, i.e. the sound horizon at the drag epoch, $r_{s}(z_{d})$, is a standard ruler which can be used to test alternative cosmologies.
To complement our SL constraints, we use the following 9 BAO points \citep[see][and references therein]{Magana:2017}to constrain the $w(z)$ functions:

\begin{itemize}
\item 6dFGS.-  $z=0.106, dz\equiv\frac{r_s(z_d)}{D_V(z)}=0.336\pm0.015$, where $D_V(z)=\frac{1}{H_0}\left[(1+z)^2D(z)^2\frac{cz}{E(z)}\right]^{1/3}$
\item WiggleZ.- $z=\left[0.44,0.6,0.73\right]$, $dz=\left[0.0870\pm0.0042, 0.0672\pm0.0031,0.0593\pm0.0020 \right]$
\item SDSS DR7 $z=0.15,0.2239\pm0.0084$ 
\item SDSS-III BOSS DR11 (a).- $z=\left[0.32,57\right]$, $dz=\left[0.1181\pm0.0022,0.0726\pm0.0007\right]$
\item SDSS-III BOSS DR11 (b).- $z=\left[2.34, 2.36\right]$, 
$\frac{D_{H}(z)}{r_s(z_d)}=\left[ 9.18 \pm 0.28, 9.00\pm0.3\right]$, where $D_{H(z)}=c/H_{0}E(z)$
\end{itemize}

It is worth noting that $r_{s}(z_{d})$ depends on the underlying  cosmology which is commonly the $\Lambda$CDM model. Moreover, the $z_{d}$ formulae employed in the BAO fitting \citep{Eisenstein:1998} were calculated for the standard cosmology. Thus, the BAO constraints could be biased due to the standard cosmology.

\subsection{Distance posteriors from CMB Planck 2015 measurements}
The information of the CMB acoustic peaks can be compressed in three quantities, their distance posteriors: the acoustic scale, $l_{A}$, the shift parameter, $R$, and the decoupling redshift, $z_{*}$. 
Several authors have proved that these quantities are almost independent of the input DE models \citep{Wang:2012}. Thus,
to constrain the $w(z)$ parameters we use the following distance posteriors for a flat $w$CDM, estimated by \citet{Neveu:2017} from Planck 2015 measurements: 
$l_{A}^{obs}=301.787\pm0.089$, $R^{obs}=1.7492\pm0.0049$, 
$z_{*}^{obs}=1089.99\pm0.29$. 

It is worth noting that the fitting formulae for these quantities \citep{Hu:1996,Bond:1997} are calculated for the standard model, however we assume that they are valid in dynamical DE models.

\subsection{SNe Ia}
Since that Type Ia Supernovae are standard candles, i.e. their light curves have the same shape after a standardization process, they have been used to measure cosmological parameters. Indeed, the apparent cosmic accelerating expansion was observed through a Hubble diagram of distant SNIa. As complementary test, we consider the compilation by \citet{Ganeshalingam:2013} which contains 586 data points of the modulus distance, $\mu$, in the redshift range $0.01<z<1.4$ which include mainly 91 points from the Lick Observatory Supernova Search (LOSS) SN Ia observations.

\subsection{H(z) measurements}
The Hubble parameter at different redshifts provide a direct measurement
of the expansion rate of the Universe. Several authors have estimated the observational Hubble data using different techniques: from clustering or BAO peaks \citep[see for instance,][]{Gaztanaga:2009} and from cosmic chronometers \citep{Jimenez:2001}. Here, we use the same sample used by \citet{Magana:2017} which contains $34$ data points in the redshift range $0.07<z<2.36$. Although some of the H(z) points were estimated from BAO data,
we assume that there is no correlation between them.
It is worth noting that the $H(z)$ points obtained from BAO could yield to biased constraints due to the underlying ($\Lambda$CDM) cosmology on $r_{s}(z_{d}).$\\

\section{Confidence Contours for the BA, FSLL I, FSLL II and SeLa parameterizations}
\label{sec:AppB}

\begin{figure}
\figurenum{B1}
\plotone{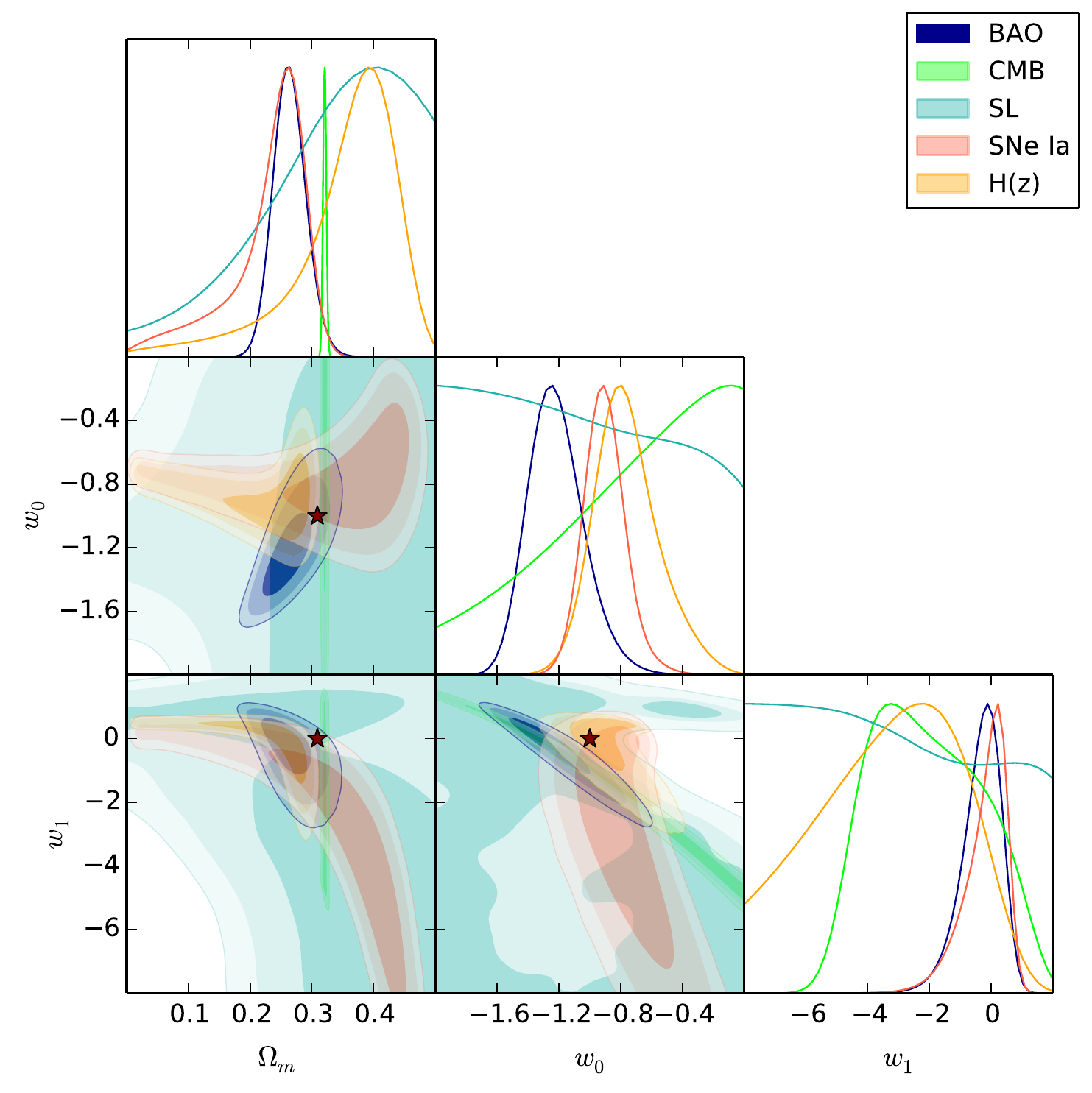}
\caption{The same as Fig. \ref{fig:JBPmodelsl10} for the BA parameterization.}
\label{fig:BAmodelsl10}
\end{figure}

\begin{figure}
\figurenum{B2}
\plotone{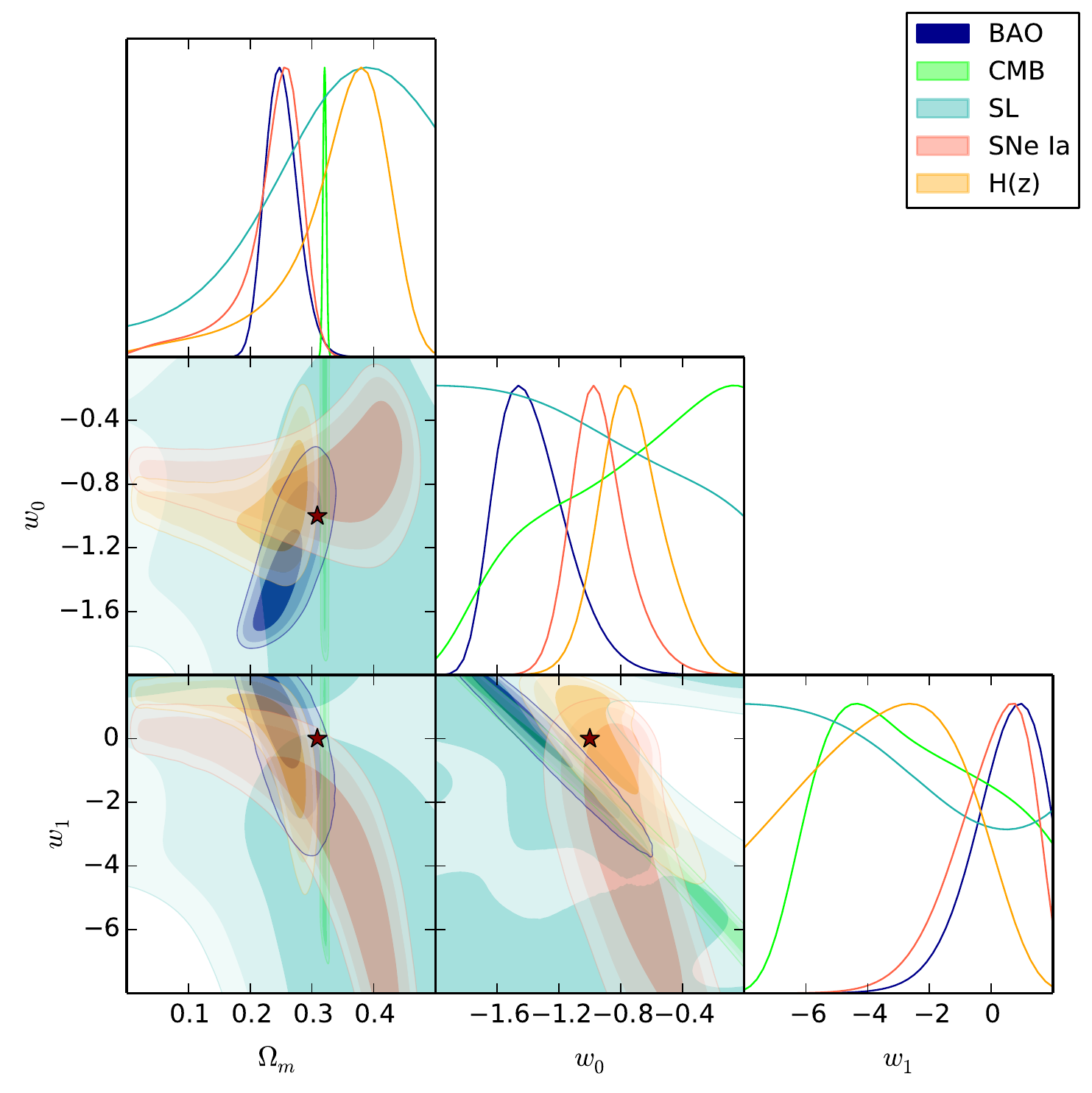}
\caption{The same as Fig. \ref{fig:JBPmodelsl10} for the FSLL I parameterization.}
\label{fig:FSLLImodelsl10}
\end{figure}

\begin{figure}
\figurenum{B3}
\plotone{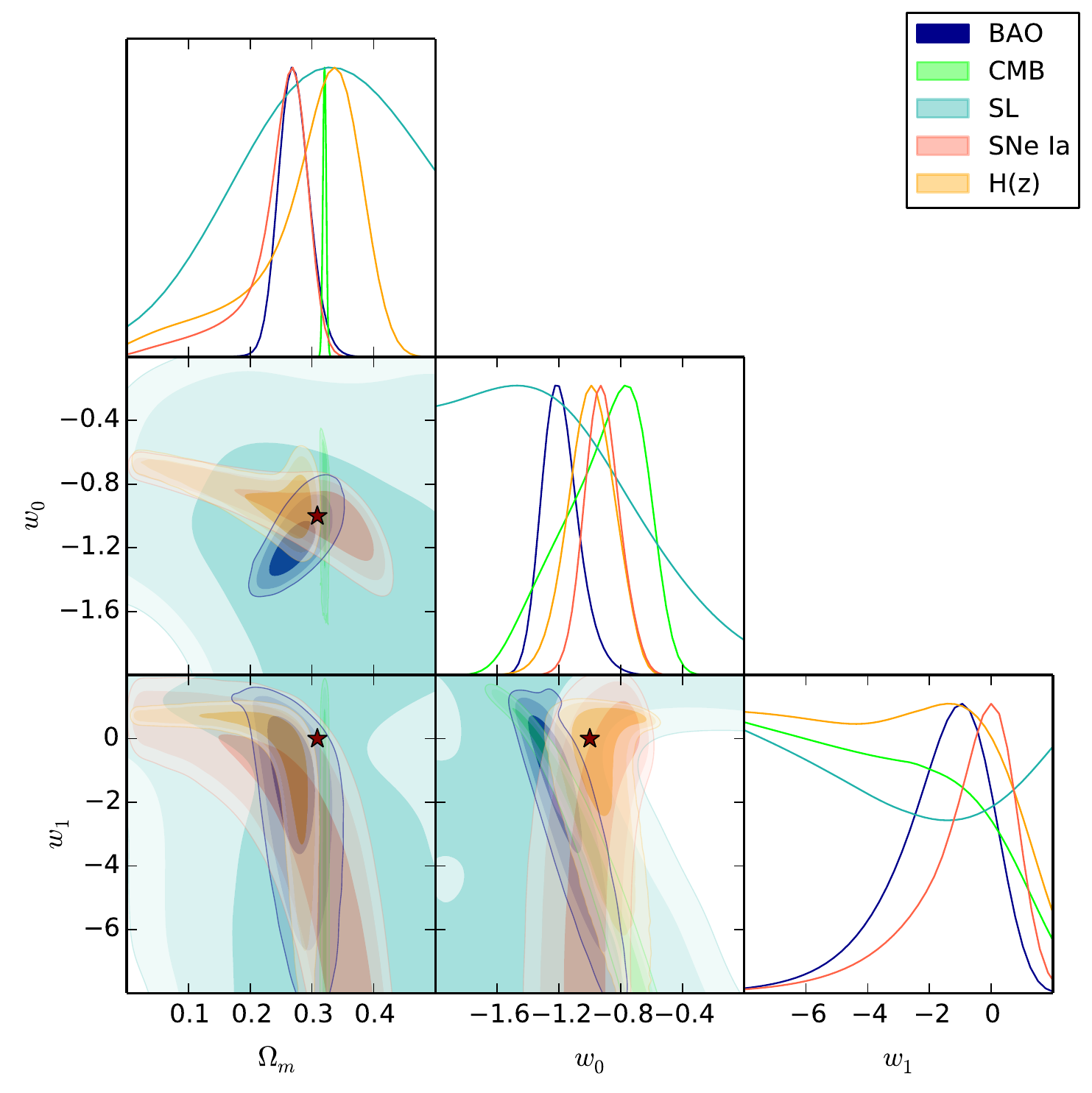}
\caption{The same as Fig. \ref{fig:JBPmodelsl10} for the FSLL II parameterization.}
\label{fig:FSLLIImodelsl10}
\end{figure}

\clearpage
\begin{figure}
\figurenum{B4}
\plotone{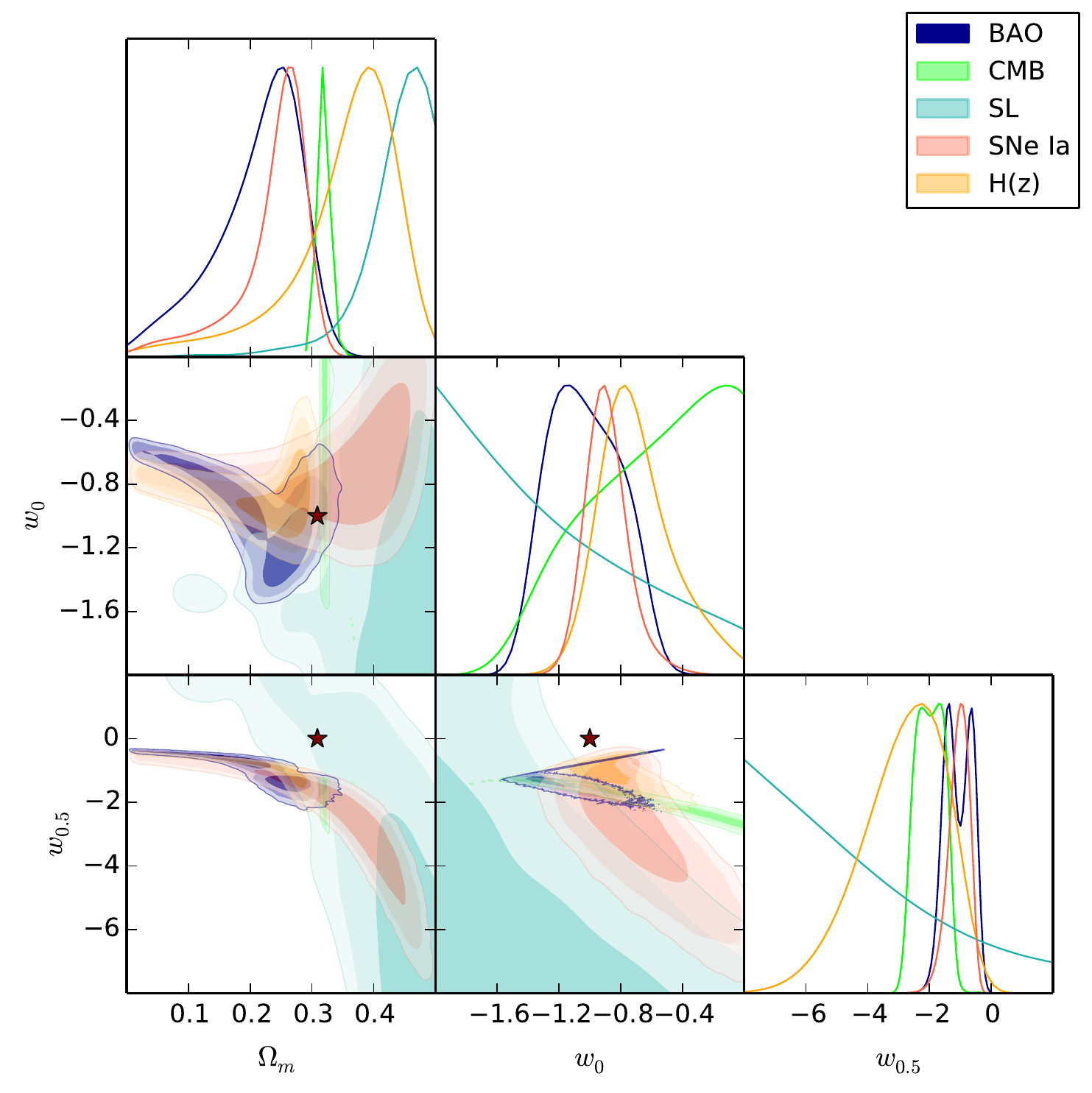}
\caption{The same as Fig. \ref{fig:JBPmodelsl10} for the SeLa parameterization.}
\label{fig:SeLamodelsl10}
\end{figure}

\bibliographystyle{aasjournal}
\bibliography{references}

\begin{thebibliography}{}
\expandafter\ifx\csname natexlab\endcsname\relax\def\natexlab#1{#1}\fi

\bibitem[{{Acebron} {et~al.}(2017){Acebron}, {Jullo}, {Limousin}, {Tilquin},
  {Giocoli}, {Jauzac}, {Malher}, \& {Richard}}]{Acebron2017}
{Acebron}, A., {Jullo}, E., {Limousin}, M., {et~al.} 2017, ArXiv e-prints,
  arXiv:1704.05380

\bibitem[{{Albrecht} {et~al.}(2006){Albrecht}, {Bernstein}, {Cahn}, {Freedman},
  {Hewitt}, {Hu}, {Huth}, {Kamionkowski}, {Kolb}, {Knox}, {Mather}, {Staggs},
  \& {Suntzeff}}]{Albrecht:2006}
{Albrecht}, A., {Bernstein}, G., {Cahn}, R., {et~al.} 2006, ArXiv Astrophysics
  e-prints, astro-ph/0609591

\bibitem[{{Barboza} \& {Alcaniz}(2008)}]{Barboza:2008}
{Barboza}, E.~M., \& {Alcaniz}, J.~S. 2008, Physics Letters B, 666, 415

\bibitem[{{Bayliss} {et~al.}(2014){Bayliss}, {Johnson}, {Gladders}, {Sharon},
  \& {Oguri}}]{Bayliss:2014}
{Bayliss}, M.~B., {Johnson}, T., {Gladders}, M.~D., {Sharon}, K., \& {Oguri},
  M. 2014, \apj, 783, 41

\bibitem[{{Bina} {et~al.}(2016){Bina}, {Pell{\'o}}, {Richard}, {Lewis},
  {Patr{\'{\i}}cio}, {Cantalupo}, {Herenz}, {Soto}, {Weilbacher}, {Bacon},
  {Vernet}, {Wisotzki}, {Cl{\'e}ment}, {Cuby}, {Lagattuta}, {Soucail}, \&
  {Verhamme}}]{Bina2016}
{Bina}, D., {Pell{\'o}}, R., {Richard}, J., {et~al.} 2016, \aap, 590, A14

\bibitem[{{Bond} {et~al.}(1997){Bond}, {Efstathiou}, \& {Tegmark}}]{Bond:1997}
{Bond}, J.~R., {Efstathiou}, G., \& {Tegmark}, M. 1997, \mnras, 291, L33

\bibitem[{{Caminha} {et~al.}(2016){Caminha}, {Grillo}, {Rosati}, {Balestra},
  {Karman}, {Lombardi}, {Mercurio}, {Nonino}, {Tozzi}, {Zitrin}, {Biviano},
  {Girardi}, {Koekemoer}, {Melchior}, {Meneghetti}, {Munari}, {Suyu}, {Umetsu},
  {Annunziatella}, {Borgani}, {Broadhurst}, {Caputi}, {Coe}, {Delgado-Correal},
  {Ettori}, {Fritz}, {Frye}, {Gobat}, {Maier}, {Monna}, {Postman}, {Sartoris},
  {Seitz}, {Vanzella}, \& {Ziegler}}]{Caminha:2016}
{Caminha}, G.~B., {Grillo}, C., {Rosati}, P., {et~al.} 2016, \aap, 587, A80

\bibitem[{{Chevallier} \& {Polarski}(2001)}]{CP:2001}
{Chevallier}, M., \& {Polarski}, D. 2001, International Journal of Modern
  Physics D, 10, 213

\bibitem[{{Chiriv{\`i}} {et~al.}(2018){Chiriv{\`i}}, {Suyu}, {Grillo},
  {Halkola}, {Balestra}, {Caminha}, {Mercurio}, \& {Rosati}}]{Chirivi:2018}
{Chiriv{\`i}}, G., {Suyu}, S.~H., {Grillo}, C., {et~al.} 2018, \aap, 614, A8

\bibitem[{{Copeland} {et~al.}(2006){Copeland}, {Sami}, \&
  {Tsujikawa}}]{Copeland:2006}
{Copeland}, E.~J., {Sami}, M., \& {Tsujikawa}, S. 2006, International Journal
  of Modern Physics D, 15, 1753

\bibitem[{{D'Aloisio} \& {Natarajan}(2011)}]{D'Aloisio:2011}
{D'Aloisio}, A., \& {Natarajan}, P. 2011, \mnras, 411, 1628

\bibitem[{{Davis}(2014)}]{Davis:2014}
{Davis}, T.~M. 2014, General Relativity and Gravitation, 46, 1731

\bibitem[{{Diego} {et~al.}(2015){Diego}, {Broadhurst}, {Benitez}, {Umetsu},
  {Coe}, {Sendra}, {Sereno}, {Izzo}, \& {Covone}}]{Diego2015}
{Diego}, J.~M., {Broadhurst}, T., {Benitez}, N., {et~al.} 2015, \mnras, 446,
  683

\bibitem[{{Eisenstein} \& {Hu}(1998)}]{Eisenstein:1998}
{Eisenstein}, D.~J., \& {Hu}, W. 1998, \apj, 496, 605

\bibitem[{{El{\'{\i}}asd{\'o}ttir} {et~al.}(2007){El{\'{\i}}asd{\'o}ttir},
  {Limousin}, {Richard}, {Hjorth}, {Kneib}, {Natarajan}, {Pedersen}, {Jullo},
  \& {Paraficz}}]{Eliasdottir:2007}
{El{\'{\i}}asd{\'o}ttir}, {\'A}., {Limousin}, M., {Richard}, J., {et~al.} 2007,
  ArXiv e-prints, arXiv:0710.5636

\bibitem[{{Feng} {et~al.}(2012){Feng}, {Shen}, {Li}, \& {Li}}]{Feng:2012}
{Feng}, C.-J., {Shen}, X.-Y., {Li}, P., \& {Li}, X.-Z. 2012, \jcap, 9, 023

\bibitem[{Ferreira {et~al.}(2017)Ferreira, Quintin, Costa, Abdalla, \&
  Wang}]{Ferreira:2017}
Ferreira, E. G.~M., Quintin, J., Costa, A.~A., Abdalla, E., \& Wang, B. 2017,
  Phys. Rev. D, 95, 043520

\bibitem[{{Foreman-Mackey} {et~al.}(2013){Foreman-Mackey}, {Hogg}, {Lang}, \&
  {Goodman}}]{Foreman-Mackey:2013}
{Foreman-Mackey}, D., {Hogg}, D.~W., {Lang}, D., \& {Goodman}, J. 2013, \pasp,
  125, 306

\bibitem[{{Ganeshalingam} {et~al.}(2013){Ganeshalingam}, {Li}, \&
  {Filippenko}}]{Ganeshalingam:2013}
{Ganeshalingam}, M., {Li}, W., \& {Filippenko}, A.~V. 2013, \mnras, 433, 2240

\bibitem[{Gaztanaga {et~al.}(2009)Gaztanaga, Cabre, \& Hui}]{Gaztanaga:2009}
Gaztanaga, E., Cabre, A., \& Hui, L. 2009, Mon. Not. Roy. Astron. Soc., 399,
  1663

\bibitem[{{Giocoli} {et~al.}(2016){Giocoli}, {Bonamigo}, {Limousin},
  {Meneghetti}, {Moscardini}, {Angulo}, {Despali}, \& {Jullo}}]{Giocoli2016}
{Giocoli}, C., {Bonamigo}, M., {Limousin}, M., {et~al.} 2016, \mnras, 462, 167

\bibitem[{{Giocoli} {et~al.}(2012){Giocoli}, {Meneghetti}, {Bartelmann},
  {Moscardini}, \& {Boldrin}}]{giocoli2012}
{Giocoli}, C., {Meneghetti}, M., {Bartelmann}, M., {Moscardini}, L., \&
  {Boldrin}, M. 2012, \mnras, 421, 3343

\bibitem[{{Golse} {et~al.}(2002){Golse}, {Kneib}, \& {Soucail}}]{Golse:2002}
{Golse}, G., {Kneib}, J.-P., \& {Soucail}, G. 2002, \aap, 387, 788

\bibitem[{{Grillo} {et~al.}(2015){Grillo}, {Suyu}, {Rosati}, {Mercurio},
  {Balestra}, {Munari}, {Nonino}, {Caminha}, {Lombardi}, {De Lucia}, {Borgani},
  {Gobat}, {Biviano}, {Girardi}, {Umetsu}, {Coe}, {Koekemoer}, {Postman},
  {Zitrin}, {Halkola}, {Broadhurst}, {Sartoris}, {Presotto}, {Annunziatella},
  {Maier}, {Fritz}, {Vanzella}, \& {Frye}}]{Grillo:2015ApJ}
{Grillo}, C., {Suyu}, S.~H., {Rosati}, P., {et~al.} 2015, \apj, 800, 38

\bibitem[{{Harvey} {et~al.}(2016){Harvey}, {Kneib}, \& {Jauzac}}]{Harvey2016}
{Harvey}, D., {Kneib}, J.~P., \& {Jauzac}, M. 2016, \mnras, 458, 660

\bibitem[{{Host}(2012)}]{Host:2012}
{Host}, O. 2012, \mnras, 420, L18

\bibitem[{{Hu} \& {Sugiyama}(1996)}]{Hu:1996}
{Hu}, W., \& {Sugiyama}, N. 1996, \apj, 471, 542

\bibitem[{{Jaroszynski} \& {Kostrzewa-Rutkowska}(2014)}]{Jaroszynski:2014}
{Jaroszynski}, M., \& {Kostrzewa-Rutkowska}, Z. 2014, \mnras, 439, 2432

\bibitem[{{Jassal} {et~al.}(2005{\natexlab{a}}){Jassal}, {Bagla}, \&
  {Padmanabhan}}]{Jassal:2005b}
{Jassal}, H.~K., {Bagla}, J.~S., \& {Padmanabhan}, T. 2005{\natexlab{a}}, \prd,
  72, 103503

\bibitem[{{Jassal} {et~al.}(2005{\natexlab{b}}){Jassal}, {Bagla}, \&
  {Padmanabhan}}]{Jassal:2005a}
---. 2005{\natexlab{b}}, \mnras, 356, L11

\bibitem[{{Jauzac} {et~al.}(2014){Jauzac}, {Cl{\'e}ment}, {Limousin},
  {Richard}, {Jullo}, {Ebeling}, {Atek}, {Kneib}, {Knowles}, {Natarajan},
  {Eckert}, {Egami}, {Massey}, \& {Rexroth}}]{Jauzac2014}
{Jauzac}, M., {Cl{\'e}ment}, B., {Limousin}, M., {et~al.} 2014, \mnras, 443,
  1549

\bibitem[{Jimenez \& Loeb(2002)}]{Jimenez:2001}
Jimenez, R., \& Loeb, A. 2002, Astrophys. J., 573, 37

\bibitem[{{Joyce} {et~al.}(2016){Joyce}, {Lombriser}, \&
  {Schmidt}}]{Joyce:2016}
{Joyce}, A., {Lombriser}, L., \& {Schmidt}, F. 2016, Annual Review of Nuclear
  and Particle Science, 66, 95

\bibitem[{{Jullo} {et~al.}(2007){Jullo}, {Kneib}, {Limousin},
  {El\'{i}asd\'{o}ttir}, {Marshall}, \& {Verdugo}}]{Jullo:2007}
{Jullo}, E., {Kneib}, J.-P., {Limousin}, M., {et~al.} 2007, \mnras, 9, 447

\bibitem[{{Jullo} {et~al.}(2010){Jullo}, {Natarajan}, {Kneib}, {D'Aloisio},
  {Limousin}, {Richard}, \& {Schimd}}]{Jullo:2010}
{Jullo}, E., {Natarajan}, P., {Kneib}, J.-P., {et~al.} 2010, Science, 329, 924

\bibitem[{{Kassiola} \& {Kovner}(1993)}]{kassiola1993}
{Kassiola}, A., \& {Kovner}, I. 1993, \apj, 417, 450

\bibitem[{{Kneib} {et~al.}(1996){Kneib}, {Ellis}, {Smail}, {Couch}, \&
  {Sharples}}]{kneib1996}
{Kneib}, J.-P., {Ellis}, R.~S., {Smail}, I., {Couch}, W.~J., \& {Sharples},
  R.~M. 1996, \apj, 471, 643

\bibitem[{Komatsu {et~al.}(2011)}]{Komatsu:2011}
Komatsu, E., {et~al.} 2011, Astrophys. J. Suppl., 192, 18

\bibitem[{{Lazkoz} {et~al.}(2005){Lazkoz}, {Nesseris}, \&
  {Perivolaropoulos}}]{Lazkoz:2005}
{Lazkoz}, R., {Nesseris}, S., \& {Perivolaropoulos}, L. 2005, \jcap, 11, 010

\bibitem[{{Li} {et~al.}(2011){Li}, {Li}, {Wang}, \& {Wang}}]{Li:2011}
{Li}, M., {Li}, X.-D., {Wang}, S., \& {Wang}, Y. 2011, Communications in
  Theoretical Physics, 56, 525

\bibitem[{{Limousin} {et~al.}(2005){Limousin}, {Kneib}, \&
  {Natarajan}}]{limousin2005}
{Limousin}, M., {Kneib}, J.-P., \& {Natarajan}, P. 2005, \mnras, 356, 309

\bibitem[{{Limousin} {et~al.}(2013){Limousin}, {Morandi}, {Sereno},
  {Meneghetti}, {Ettori}, {Bartelmann}, \& {Verdugo}}]{Limousin2013}
{Limousin}, M., {Morandi}, A., {Sereno}, M., {et~al.} 2013, \ssr, 177, 155

\bibitem[{{Limousin} {et~al.}(2007){Limousin}, {Richard}, {Jullo}, {Kneib},
  {Fort}, {Soucail}, {El{\'{\i}}asd{\'o}ttir}, {Natarajan}, {Ellis}, {Smail},
  {Czoske}, {Smith}, {Hudelot}, {Bardeau}, {Ebeling}, {Egami}, \&
  {Knudsen}}]{Limousin:2007ApJ}
{Limousin}, M., {Richard}, J., {Jullo}, E., {et~al.} 2007, \apj, 668, 643

\bibitem[{{Limousin} {et~al.}(2010){Limousin}, {Jullo}, {Richard}, {Cabanac},
  {Suyu}, {Halkola}, {Kneib}, {Gavazzi}, \& {Soucail}}]{Limousin2010}
{Limousin}, M., {Jullo}, E., {Richard}, J., {et~al.} 2010, \aap, 524, A95

\bibitem[{{Limousin} {et~al.}(2016){Limousin}, {Richard}, {Jullo}, {Jauzac},
  {Ebeling}, {Bonamigo}, {Alavi}, {Cl{\'e}ment}, {Giocoli}, {Kneib}, {Verdugo},
  {Natarajan}, {Siana}, {Atek}, \& {Rexroth}}]{Limousin:2016}
{Limousin}, M., {Richard}, J., {Jullo}, E., {et~al.} 2016, \aap, 588, A99

\bibitem[{{Linder}(2003)}]{Linder:2003}
{Linder}, E.~V. 2003, Physical Review Letters, 90, 091301

\bibitem[{{Link} \& {Pierce}(1998)}]{Link:1998ApJ}
{Link}, R., \& {Pierce}, M.~J. 1998, \apj, 502, 63

\bibitem[{Maga\~na {et~al.}(2014)Maga\~na, C\'ardenas, \& Motta}]{Magana:2014}
Maga\~na, J., C\'ardenas, V.~H., \& Motta, V. 2014, JCAP, 1410, 017

\bibitem[{{Maga{\~n}a} {et~al.}(2017){Maga{\~n}a}, {Motta}, {C{\'a}rdenas}, \&
  {Fo{\"e}x}}]{Magana:2017}
{Maga{\~n}a}, J., {Motta}, V., {C{\'a}rdenas}, V.~H., \& {Fo{\"e}x}, G. 2017,
  \mnras, 469, 47

\bibitem[{Maga\~na {et~al.}(2015)Maga\~na, Motta, Cardenas, Verdugo, \&
  Jullo}]{Magana:2015}
Maga\~na, J., Motta, V., Cardenas, V.~H., Verdugo, T., \& Jullo, E. 2015,
  Astrophys. J., 813, 69

\bibitem[{{McCully} {et~al.}(2014){McCully}, {Keeton}, {Wong}, \&
  {Zabludoff}}]{McCully:2014}
{McCully}, C., {Keeton}, C.~R., {Wong}, K.~C., \& {Zabludoff}, A.~I. 2014,
  \mnras, 443, 3631

\bibitem[{{McCully} {et~al.}(2017){McCully}, {Keeton}, {Wong}, \&
  {Zabludoff}}]{McCully2017}
---. 2017, \apj, 836, 141

\bibitem[{{Meneghetti} {et~al.}(2016){Meneghetti}, {Natarajan}, {Coe},
  {Contini}, {De Lucia}, {Giocoli}, {Acebron}, {Borgani}, {Bradac}, {Diego},
  {Hoag}, {Ishigaki}, {Johnson}, {Jullo}, {Kawamata}, {Lam}, {Limousin},
  {Liesenborgs}, {Oguri}, {Sebesta}, {Sharon}, {Williams}, \&
  {Zitrin}}]{meneghetti2016}
{Meneghetti}, M., {Natarajan}, P., {Coe}, D., {et~al.} 2016, ArXiv e-prints,
  arXiv:1606.04548

\bibitem[{{Miralda-Escude} \& {Babul}(1995)}]{Miralda1995}
{Miralda-Escude}, J., \& {Babul}, A. 1995, \apj, 449, 18

\bibitem[{{Monna} {et~al.}(2017){Monna}, {Seitz}, {Balestra}, {Rosati},
  {Grillo}, {Halkola}, {Suyu}, {Coe}, {Caminha}, {Frye}, {Koekemoer},
  {Mercurio}, {Nonino}, {Postman}, \& {Zitrin}}]{Monna2017}
{Monna}, A., {Seitz}, S., {Balestra}, I., {et~al.} 2017, \mnras,
  arXiv:1605.08784

\bibitem[{{Mortonson} {et~al.}(2014){Mortonson}, {Weinberg}, \&
  {White}}]{Mortonson:2014}
{Mortonson}, M.~J., {Weinberg}, D.~H., \& {White}, M. 2014, ArXiv e-prints,
  arXiv:1401.0046

\bibitem[{{Neveu} {et~al.}(2017){Neveu}, {Ruhlmann-Kleider}, {Astier},
  {Besan{\c c}on}, {Guy}, {M{\"o}ller}, \& {Babichev}}]{Neveu:2017}
{Neveu}, J., {Ruhlmann-Kleider}, V., {Astier}, P., {et~al.} 2017, \aap, 600,
  A40

\bibitem[{{Pantazis} {et~al.}(2016){Pantazis}, {Nesseris}, \&
  {Perivolaropoulos}}]{Pantazis:2016}
{Pantazis}, G., {Nesseris}, S., \& {Perivolaropoulos}, L. 2016, \prd, 93,
  103503

\bibitem[{Perlmutter {et~al.}(1999)Perlmutter, Aldering, Goldhaber, Knop,
  Nugent, others, \& Project}]{Perlmutter:1999}
Perlmutter, S., Aldering, G., Goldhaber, G., {et~al.} 1999, The Astrophysical
  Journal, 517, 565

\bibitem[{{Planck Collaboration} {et~al.}(2016{\natexlab{a}}){Planck
  Collaboration}, {Ade}, {Aghanim}, {Arnaud}, {Ashdown}, {Aumont},
  {Baccigalupi}, {Banday}, {Barreiro}, {Bartlett}, \& et~al.}]{Planck:2015XIII}
{Planck Collaboration}, {Ade}, P.~A.~R., {Aghanim}, N., {et~al.}
  2016{\natexlab{a}}, \aap, 594, A13

\bibitem[{{Planck Collaboration} {et~al.}(2016{\natexlab{b}}){Planck
  Collaboration}, {Ade}, {Aghanim}, {Arnaud}, {Ashdown}, {Aumont},
  {Baccigalupi}, {Banday}, {Barreiro}, {Bartolo}, \& et~al.}]{Planck:2015XIV}
---. 2016{\natexlab{b}}, \aap, 594, A14

\bibitem[{Riess {et~al.}(1998)Riess, Filippenko, Challis, Clocchiatti, Diercks,
  {et~al.}}]{Riess:1998}
Riess, A.~G., Filippenko, A.~V., Challis, P., {et~al.} 1998, The Astronomical
  Journal, 116, 1009

\bibitem[{Salvatelli {et~al.}(2014)Salvatelli, Said, Bruni, Melchiorri, \&
  Wands}]{Salvatelli:2014}
Salvatelli, V., Said, N., Bruni, M., Melchiorri, A., \& Wands, D. 2014, Phys.
  Rev. Lett., 113, 181301

\bibitem[{{Sendra} \& {Lazkoz}(2012)}]{Sendra:2012}
{Sendra}, I., \& {Lazkoz}, R. 2012, \mnras, 422, 776

\bibitem[{{Soucail} {et~al.}(2004){Soucail}, {Kneib}, \&
  {Golse}}]{Soucail:2004}
{Soucail}, G., {Kneib}, J.-P., \& {Golse}, G. 2004, \aap, 417, L33

\bibitem[{{Tu} {et~al.}(2008){Tu}, {Limousin}, {Fort}, {Shu}, {Sygnet},
  {Jullo}, {Kneib}, \& {Richard}}]{Tu2008}
{Tu}, H., {Limousin}, M., {Fort}, B., {et~al.} 2008, \mnras, 386, 1169

\bibitem[{{Umetsu} {et~al.}(2015){Umetsu}, {Sereno}, {Medezinski}, {Nonino},
  {Mroczkowski}, {Diego}, {Ettori}, {Okabe}, {Broadhurst}, \&
  {Lemze}}]{Umetsu2015}
{Umetsu}, K., {Sereno}, M., {Medezinski}, E., {et~al.} 2015, \apj, 806, 207

\bibitem[{{Wang} {et~al.}(2006){Wang}, {Li}, {Kauffmann}, \& {De
  Lucia}}]{wang2006}
{Wang}, L., {Li}, C., {Kauffmann}, G., \& {De Lucia}, G. 2006, \mnras, 371, 537

\bibitem[{{Wang} {et~al.}(2016){Wang}, {Hu}, {Li}, \& {Li}}]{Wang:2016}
{Wang}, S., {Hu}, Y., {Li}, M., \& {Li}, N. 2016, \apj, 821, 60

\bibitem[{{Wang}(2008)}]{Wang:2008}
{Wang}, Y. 2008, \prd, 77, 123525

\bibitem[{{Wang} {et~al.}(2012){Wang}, {Chuang}, \& {Mukherjee}}]{Wang:2012}
{Wang}, Y., {Chuang}, C.-H., \& {Mukherjee}, P. 2012, \prd, 85, 023517

\bibitem[{Weinberg(1989)}]{Weinberg}
Weinberg, S. 1989, Reviews of Modern Physics, 61

\bibitem[{Zeldovich(1968)}]{Zeldovich}
Zeldovich, Y.~B. 1968, Soviet Physics Uspekhi, 11

\bibitem[{{Zhao} {et~al.}(2017){Zhao}, {Raveri}, {Pogosian}, {Wang},
  {Crittenden}, {Handley}, {Percival}, {Beutler}, {Brinkmann}, {Chuang},
  {Cuesta}, {Eisenstein}, {Kitaura}, {Koyama}, {L'Huillier}, {Nichol}, {Pieri},
  {Rodriguez-Torres}, {Ross}, {Rossi}, {S{\'a}nchez}, {Shafieloo}, {Tinker},
  {Tojeiro}, {Vazquez}, \& {Zhang}}]{Zhao:2017Nat}
{Zhao}, G.-B., {Raveri}, M., {Pogosian}, L., {et~al.} 2017, Nature Astronomy,
  1, 627

\bibitem[{Zitrin {et~al.}(2012)Zitrin, Rosati, Nonino, Grillo, Postman, Coe,
  Seitz, Eichner, Broadhurst, Jouvel, Balestra, Mercurio, Scodeggio, Benítez,
  Bradley, Ford, Host, Jimenez-Teja, Koekemoer, Zheng, Bartelmann, Bouwens,
  Czoske, Donahue, Graur, Graves, Infante, Jha, Kelson, Lahav, Lazkoz, Lemze,
  Lombardi, Maoz, McCully, Medezinski, Melchior, Meneghetti, Merten, Molino,
  Moustakas, Ogaz, Patel, Regoes, Riess, Rodney, Umetsu, \& der
  Wel}]{Zitrin:2012}
Zitrin, A., Rosati, P., Nonino, M., {et~al.} 2012, The Astrophysical Journal,
  749, 97

\bibitem[{{Zitrin} {et~al.}(2015){Zitrin}, {Fabris}, {Merten}, {Melchior},
  {Meneghetti}, {Koekemoer}, {Coe}, {Maturi}, {Bartelmann}, {Postman},
  {Umetsu}, {Seidel}, {Sendra}, {Broadhurst}, {Balestra}, {Biviano}, {Grillo},
  {Mercurio}, {Nonino}, {Rosati}, {Bradley}, {Carrasco}, {Donahue}, {Ford},
  {Frye}, \& {Moustakas}}]{Zitrin2015}
{Zitrin}, A., {Fabris}, A., {Merten}, J., {et~al.} 2015, \apj, 801, 44

\end{thebibliography}

\end{document}